\documentclass[acmsmall]{acmart}

\usepackage[utf8]{inputenc} 
\usepackage[T1]{fontenc}    
\usepackage{hyperref}       
\usepackage{url}            
\usepackage{booktabs}       
\usepackage{amsfonts}       
\usepackage{nicefrac}       
\usepackage{microtype}      
\usepackage{lipsum}

\usepackage{mathptmx} 
\usepackage[title]{appendix}
\usepackage{graphicx}
\usepackage{amsmath}

\usepackage{amssymb}
\usepackage{epstopdf}
\usepackage{multirow}
\usepackage{algorithm}
\usepackage{algpseudocode}
\usepackage{hhline}
\usepackage{enumitem}
\usepackage{listings}
\usepackage{anyfontsize}
\usepackage{wrapfig}
\usepackage{footnote}
\usepackage{tablefootnote}
\usepackage{xcolor}
\usepackage{wrapfig}
\usepackage{soul}
\makesavenoteenv{tabular}
\makesavenoteenv{table}
\graphicspath{ {figures/} }

\usepackage{tikz}
\newcommand*\circled[1]{\tikz[baseline=(char.base)]{%
            \node[shape=circle,draw,inner sep=0.5pt] (char) {#1};}}

\lstdefinestyle{myStyle}{
  language=SQL,
  basicstyle=\small\ttfamily,
  moredelim=[is][\underbar]{_}{_},
  keepspaces=true
}

\definecolor{comment}{HTML}{3F7F7F}
\makeatletter
\def\hlinewd#1{%
\noalign{\ifnum0=`}\fi\hrule \@height #1 \futurelet
\reserved@a\@xhline}
\makeatother

\usepackage{array}
\newcolumntype{L}[1]{>{\raggedright\arraybackslash}p{#1}}
\newcolumntype{C}[1]{>{\centering\arraybackslash}p{#1}}
\newcolumntype{R}[1]{>{\raggedleft\arraybackslash}p{#1}}

\newenvironment{nohyphen}
 {\par\hyphenpenalty=10000}
 {\par}

\newcommand{\mysize}{\fontsize{7.4pt}{2pt}\selectfont}

\usepackage{xcolor,colortbl}

\newcolumntype{y}{>{\columncolor{yellow}}c}

\usepackage{fancyhdr}
\usepackage[normalem]{ulem}

\AtBeginDocument{%
  \providecommand\BibTeX{{%
    \normalfont B\kern-0.5em{\scshape i\kern-0.25em b}\kern-0.8em\TeX}}}

\setcopyright{acmcopyright}
\acmJournal{TODAES}
\acmYear{2021} \acmVolume{27} \acmNumber{3} \acmArticle{25}
\acmMonth{11} \acmPrice{15.00}\acmDOI{10.1145/3490176}

\begin{document}

\title{Toward Taming the Overhead Monster for Data-Flow Integrity}

\author{Lang Feng}
\email{flang@nju.edu.cn}

\affiliation{%
  \institution{Nanjing University}
  \streetaddress{163 Xianlin Road}
  \city{Qixia District}
  \state{Nanjing}
  \country{China}
  \postcode{210023}
}

\affiliation{%
  \institution{Texas A\&M University}
  \streetaddress{400 Bizzell Street}
  \city{College Station}
  \state{Texas}
  \country{USA}
  \postcode{77840}
}

\author{Jiayi Huang}
\email{jyhuang@ucsb.edu}

\affiliation{%
  \institution{University of California, Santa Barbara}
  \state{California}
  \country{USA}
  \postcode{93106}
}

\author{Jeff Huang}
\email{jeffhuang@tamu.edu}

\author{Jiang Hu}
\email{jianghu@tamu.edu}

\affiliation{%
  \institution{Texas A\&M University}
  \streetaddress{400 Bizzell Street}
  \city{College Station}
  \state{Texas}
  \country{USA}
  \postcode{77840}
}


\begin{abstract}
Data-Flow Integrity (DFI) is a well-known approach to effectively detecting a wide range of software attacks.
However, its real-world application has been quite limited so far because of the prohibitive performance overhead it incurs.
Moreover, the overhead is enormously difficult to overcome without substantially lowering the DFI criterion. In this work, an analysis is performed to understand the main factors contributing to the overhead. 
Accordingly, a hardware-assisted parallel approach is proposed to 
tackle the overhead challenge.
Simulations on SPEC CPU 2006 benchmark show that the proposed approach can completely enforce the DFI defined in the original seminal work
while reducing performance overhead by $4\times$, on average.
\end{abstract}

\ccsdesc[500]{Computer systems organization~Architectures}
\ccsdesc[500]{Security and privacy~Systems security}
\keywords{Data-Flow Integrity, Processing in Memory}

\maketitle
\newcommand\blfootnote[1]{%
\begingroup
\renewcommand\thefootnote{}\footnote{#1}%
\addtocounter{footnote}{-1}%
\endgroup
}
\blfootnote{This work is partially supported by NSF (CNS-1618824) and NSF (CCF-1815583).}

\vspace*{-2mm}
\section{Introduction}

\begin{sloppypar}
Data-Flow Integrity (DFI) is a regulation to ensure that data to be
accessed are written by legitimate instructions~\cite{Castro06}. As such, DFI enforcement can identify unwanted data modifications that are not consistent with the programmer's intention. 
It can detect a wide variety of security attacks including control data attacks 
such as Jump-Oriented Programming (JOP)~\cite{jop} and Return-Oriented Programming (ROP)~\cite{rop}, and non-control data attacks such as Heartbleed~\cite{heartbleed} and the heap overflow attack to  Nullhttpd~\cite{nullhttpd}. 
As a large number of software attacks rely on data modifications, DFI is a single principle that is effective for many 
different attack scenarios including future potential ones.
In fact, its defense scope is a much bigger superset of 
Control-Flow Integrity (CFI){~\cite{Abadiccs05}}, which is another well-known software security approach.
\end{sloppypar}

The concept of DFI was introduced in 2006 by the seminal work~\cite{Castro06} and has received a great amount of attention thereafter due to its potential of being a powerful security measure.
To differentiate from the approaches in terms of granularity, the DFI enforcement in the seminal work{~\cite{Castro06}} is named \textbf{complete DFI} in this paper.
However, a complete DFI enforcement as in~\cite{Castro06} incurs more than $100\%$ performance overhead even though several optimization techniques have been applied. 
Indeed, the huge overhead seems
inevitable as every data access needs to be examined. Due to this intrinsic
difficulty, there have been few follow-up works on DFI despite its widely recognized importance. 
This is in sharp contrast to CFI~\cite{Abadiccs05}, which has much more published studies~\cite{Lee17,Guoscw14,Geasplos17,Liuhpca17,XiaDSN12,Davi12mocfi,Hu18}.

The few later works on DFI~\cite{Song16,Song16_2,Akritidis08,Liu18_2} reduce the overhead by exploiting partial DFI, whose criteria are substantially lower than the original DFI definition{~\cite{Castro06}}.
The Hardware-Assisted Data-Flow Isolation (HDFI){~\cite{Song16}} is one example. It partitions data into two regions, and only requires that data to be read and written
must be consistent in the same region. In other words, it reports
a violation only when data intends to be in one region but is actually written by an instruction for another region. Although its overhead is very small, the enforcement granularity is very coarse 
and may miss attacks that mingle different data within the same region,
By contrast, a complete DFI{~\cite{Castro06}} can isolate data among dozens of thousands of regions, i.e., a resolution
$> 30000\times$ higher than HDFI.
Therefore, the security price that HDFI paid for its overhead reduction can be very high.

Enforcing the complete DFI~\cite{Castro06} with practically acceptable overhead is a huge challenge. 
Different from most of existing overhead reduction techniques~\cite{Song16,Song16_2,Akritidis08,Liu18_2}, which rely on 
lowering the DFI criterion, we pursue a new approach that 
exploits additional hardware while the original DFI~\cite{Castro06}
can still be completely enforced.
As hardware cost becomes increasingly affordable along with the progress of semiconductor technology, reducing performance
overhead at the expense of extra hardware is a promising direction.

We first conduct an extensive performance analysis of DFI. Surprisingly, the frequent DFI data access does not lead to frequent memory access and thus, memory access is not a bottleneck,
but the other computations involved in DFI enforcement contribute the most to the overhead.
We propose a parallel and asynchronous approach, where
most of the DFI computations are performed in another processor core. However, a straightforward 
software-based parallel computing still experiences 
huge overhead resulted from runtime information collection
and communications with the other processor core. 
Therefore, we develop a new hardware technique to further trim down  
the overhead. 
This hardware-assisted parallel approach also includes
new software instrumentation techniques,
lossless data compression and runtime optimization techniques.
For the ease of deployment, we intend to minimize the dependence on computing infrastructure changes.
Except the necessary circuits and software instrumentation, 
our approach does not rely on using new instructions, or OS modifications.

Overall, the proposed approach reduces performance overhead 
from $161\%$ of \cite{Castro06}
to an average of $36\%$ on the same SPEC CPU 2006 benchmarks. As it is a complete DFI enforcement,
it can detect a wide range of security attacks and cover cases that cannot 
be handled by the previous low-overhead methods~\cite{Song16,Song16_2,Akritidis08,Liu18_2}. 
Our approach provides a solution with a security-overhead tradeoff in complement to
existing methods~\cite{Song16,Song16_2,Akritidis08,Liu18_2}. 
A brief comparison with existing methods is summarized in
Table~\ref{tab:work_comp}. The contributions of this work are as follows. 
\begin{itemize}[leftmargin=1em]
\setlength\itemsep{0em}
\item An overhead breakdown analysis is performed to understand the main performance bottlenecks in software DFI.
\item This is the first hardware approach for complete DFI enforcement, to the best of our knowledge.
\item Two variants of the proposed approach are investigated, one for Processing-In-Memory (PIM) and the other for Chip Multiprocessor (CMP).
\item The tradeoff between DFI violation detection latency and performance overhead is studied.
\item Our approach achieves about $4\times$ overhead reduction, which is a major progress for complete DFI since 2006.
\end{itemize}

The rest of this paper is organized as follows.
Section{~\ref{sec:background}} introduces the background. The threat model and system assumptions are introduced in Section{~\ref{sec:theatmodel}}. The related work is briefly reviewed in Section{~\ref{sec:work}}, and characterizations are performed in Section{~\ref{sec:sys}} to uncover the key factor for the performance overhead.
Section{~\ref{sec:overview}} provides an
overview of our approach. Next, Sections {\ref{sec:instru}}, {\ref{sec:hardware}} and {\ref{sec:checkprog}} describe the three critical parts of our design.
The experiment results are shown in Section{~\ref{sec:exp}}. Section{~\ref{sec:limit}} discusses the trade-off of our approach and broader applications.
Finally, we conclude in Section{~\ref{sec:conclu}}.

\begin{table}[!htb]
\scriptsize
\def\arraystretch{1.4}
\centering
\caption{Comparison between our work and others.}
\label{tab:work_comp}
    \begin{tabular}{cccccc}
    \hlinewd{0.6pt}
       \multirow{3}{1.2cm}{\centering \textbf{Method\footnote{All the listed works need code static analysis and instrumentation.}}} & \multirow{3}{1.3cm}{\centering \textbf{Performance Overhead}} & \multirow{3}{1.6cm}{\centering \textbf{DFI Enforcement Completeness}} & \multirow{3}{*}{\textbf{Approach}} & \multirow{3}{0.9cm}{\centering \textbf{New Instruction}} & \multirow{3}{0.9cm}{\centering \textbf{OS Change}} \\ 
       & & & & & \\
       & & & & & \\ \hline
SW DFI~\cite{Castro06} & 161\% & Complete & SW & $\times$ & $\times$ \\
KENALI~\cite{Song16_2} & 7-15\% & Partial & SW  & $\times$ & $\surd$ \\
WIT~\cite{Akritidis08} & 7\% & Partial & SW & $\times$ & $\times$ \\
CHERI~\cite{Watson15} & 5-20\% & Partial & HW  & $\surd$ & $\surd$ \\
TMDFI~\cite{Liu18_2} & 39\% & Partial& HW & $\surd$ & $\times$  \\
HDFI~\cite{Song16} & $<$2\% & Partial& HW & $\surd$ & $\surd$  \\
\textbf{Our work} & \textbf{36\%} & \textbf{Complete} & \textbf{HW} & $\times$ & $\times$ \\\hlinewd{0.6pt}
    \end{tabular}

\end{table}

\vspace*{-2mm}
\section{Background on Data-Flow Integrity}
\label{sec:background}

Data-flow integrity requires that data to be loaded from memory can only be stored by 
legitimate instructions that are consistent with the programmer's original intention~\cite{Castro06}. 
Every instruction in a program is assigned a numerical \textbf{identifier} through automatic code instrumentation. 
If the data loaded by instruction \texttt{A} was most recently stored by instruction \texttt{B}, the \textbf{reaching definition} of \texttt{A} is \texttt{B}, and is represented by the identifier of \texttt{B}.
Each instruction that can load data from memory has its own \textbf{Reaching Definition Set (RDS)}, which consists of all the allowed reaching definitions of this instruction.
A static software analysis can be performed for a program to obtain the RDSs for all relevant instructions. 
In the example of Fig.~\ref{fig:dfi_ex2}, ``\texttt{store x y}'' means storing variable \texttt{x} at address \texttt{y}, 
``\texttt{load x y}'' is to load the data at address \texttt{y} to variable \texttt{x}, ``\texttt{cmp x y}'' is to compare the values of variable \texttt{x} and variable \texttt{y}, and ``\texttt{jne label}'' implies a conditional branch to the location marked by \texttt{label}, if the values in the previous comparison are different. If the identifier of each instruction is the same as its line number, the RDS of instruction ``\texttt{load x3 addr1}'' (line 6) is \{5\}, and the RDS of instruction ``\texttt{load x4 addr1}'' (line 8) is \{1, 5\}.
DFI requires that all the instructions that can load data from memory are consistent with their RDSs, i.e., when executing an instruction \texttt{A} that loads data from memory, the data should be indeed most recently stored by one of the instructions in the RDS of \texttt{A}.  Hence, the identifier of the latest instruction that stores a data needs
to be tracked for the data. Such identifiers for all data form 
a \textbf{Reaching Definition Table (RDT)}. 

In summary, given a program, the information required for DFI enforcement and their locations are as follows.
\begin{enumerate}[itemsep=0ex,leftmargin=1.5em] 
\item RDS (Reaching Definition Set) for all \texttt{load} instructions in the program. This information never changes throughout the program execution, and it can be loaded into the memory once in the beginning. 
\item RDT (Reaching Definition Table). This information changes dynamically during a program execution. It is stored in the memory and maintained by the computing resource for DFI enforcement.
\item Target instruction information. A  \textbf{target instruction} is an instruction in the program to be enforced for DFI.
Mainly two types of instructions are involved: \texttt{load} instructions for which DFI enforcement is performed, and \texttt{store} instructions that affect RDT. These two pieces of information change at runtime and need to be obtained by the computing resource for DFI enforcement. It consists of the following components:
  \begin{itemize}[leftmargin=1em]
  \item Instruction identifier.
  \item Instruction type: either \texttt{load} or \texttt{store}.
  \item Target address of \texttt{load} or \texttt{store}.
  \end{itemize}
\end{enumerate}

After all the three kinds of information are obtained, DFI enforcement can be performed.

\begin{figure}[!hbt]
\begin{lstlisting}[basicstyle=\ttfamily\scriptsize, xleftmargin=3em,numbers=left]
store x1 addr1
store x2 addr2
cmp x1 x2
jne label
store x2 addr1
load x3 addr1     // RDS: {5}
label:
load x4 addr1     // RDS: {1, 5}
\end{lstlisting}
    \vspace{-1.5em}
    \centering
    \caption{A code example for illustrating DFI, where the line numbers are used as identifiers (IDs) for the corresponding instructions for simplicity.}
    \label{fig:dfi_ex2}
\vspace{-2ex}
\end{figure}

\section{Threat Model and System Assumptions}
\label{sec:theatmodel}

Following the typical threat model of most related work, it is assumed that the attackers are able to leverage the possible software vulnerabilities to corrupt any locations in the memory. Once the attack is successfully performed, the attackers can take any desired actions. 
The software vulnerabilities may exist in any places of the user programs. 
Note that any attacks that leverage hardware vulnerabilities are not considered in our threat model. For example, rowhammer{~\cite{kim2014rowhammer}} and cache side-channel attacks{~\cite{Kocher2018spectre}} are out of the scope. Meanwhile, the attacks that maliciously modify the instructions can be simply protected by Write XOR Execute (W$\oplus$X) technique{~\cite{wx}}, and they are not included in this work.

For our system, we assume that all the software can be static analyzed, and the static analysis is assumed to be accurate. The static analysis tool can provide the software programs' RDSs of all the instructions that can load data. The DFI software toolchain and the hardware of our system are assumed to be trusted and bug free.

Under this threat model, DFI is a superset of Control-Flow Integrity (CFI){~\cite{Abadiccs05}}, which only regulates instruction flow transitions toward target addresses conforming to the original design intention. Attackers have to modify the control data, such as the target address for an indirect
branch, to change the control flow. By protecting all the data, DFI can also prevent control-flow attacks. Additionally, DFI can protect non-control data that cannot be covered by CFI.

\section{Previous Work}
\label{sec:work}

The concept of Data-Flow Integrity (DFI) was proposed in 
the seminal work~\cite{Castro06} in 2006. This work also provides a software implementation technique 
and optimization techniques for overhead reduction.  
Although the DFI enforcement procedure is simple, its performance overhead 
is intrinsically huge as the enforcement needs to be 
conducted for tremendous data.




The few later previous works~\cite{Song16_2,Akritidis08,Song16,Watson15,Liu18_2}
achieved much lower overhead by focusing on partial DFI. 
The work of \cite{Song16_2} is restricted to only certain selected data for kernel software.
One of its main contributions is the techniques on how to select data to be protected. 
Although its performance overhead is only $7-15\%$, its application is restrictive and 
misses many attacks at user programs. For example, Nullhttpd~\cite{nullhttpd}, Heartbleed~\cite{heartbleed} and data-oriented programming~\cite{Hu16} are conducted at user level and thus not handled by this technique.
By contrast, our approach covers both kernel and user level programs.

While DFI involves both \texttt{load} and \texttt{store} instructions, the scope of
Write Integrity Testing (WIT)~\cite{Akritidis08} is restricted to \texttt{store}. 
It requires that each \texttt{store} instruction can only write to certain data objects, 
and each indirect call can only call certain functions.
Although its overhead is at most 25\%, it does not cover
\texttt{load} instructions. Thus, an unsafe \texttt{load} instruction may read more bytes than 
the programmer's intention, and consequently information leak may occur, e.g.,
Heartbleed~\cite{heartbleed} is an attack that WIT would fail to detect. 
Another related work HardScope{~\cite{Nyman17}} also restricts the memory access behaviors of each function with different memory access rules, which only allow certain data to be accessed. Thus, it prevents the instructions in the unprivileged functions from accessing the privileged data. While HardScope has low performance overhead, it does not distinguish each \texttt{store} and \texttt{load}, thus, it is less fine-grained than complete DFI.

Data isolation is another approach to protecting data with relatively low overhead. 
A hardware solution for data-flow isolation, called HDFI, is proposed in \cite{Song16}. It designates
two data regions, a sensitive one and a non-sensitive one. A 1-bit tag is
employed to tell the region that a data belongs to. Instruction set is modified such that
the tags can be read and set. Moreover, the processor hardware and the operating system 
 also need changes. If data belongs to one region, it cannot be written by 
an instruction for the other region. Although the isolation 
helps security, it cannot handle the case where \texttt{load/store} instructions for
different data of the same region are mingled. 
Consider the example in Fig.{~\ref{fig:hdfi_ex}}, where input data
are first written into \texttt{u0} and \texttt{u1} at lines 10 and 11. Later, the data are copied to buffers at lines 13-15.
If there is buffer overflow when executing line 10, i.e.,
the input data size exceeds 256, then offset \texttt{u0-$>$off} is modified unintentionally. 
Then, line 13 may copy \texttt{user0}'s data to other users' buffers through the modified \texttt{u0-$>$off}. 
Meanwhile, \texttt{user1} can write to \texttt{user2}'s buffer at line 14 in the same way. 
As HDFI partitions data into only two regions, one of the user pairs --- 
(\texttt{user0}, \texttt{user1}), (\texttt{user0}, \texttt{user2}) or (\texttt{user1}, \texttt{user2}) must share the same region. 
Consequently, the former user in a pair can attack the latter in the pair without being detected by HDFI.
By contrast, to a certain degree,
the original DFI{~\cite{Castro06}} can be regarded as data isolation
among individual instructions. If 16 bits are used for each instruction 
identifier, it is equivalent to isolation among up to $2^{16}$ regions.
Compared to the only 2 regions of HDFI{~\cite{Song16}}, the resolution
of the original DFI is $2^{15} = 32768$ times higher.
Thus, its 
low overhead of $<2\%$ comes with the price of very 
coarse-grained security resolution.
Similar to HDFI, TMDFI~\cite{Liu18_2} also enforces DFI by a tag-based approach, and it results in 39\% overhead. However, TMDFI only uses 8 bits for the tag, and can only isolate $2^{8} = 256$ regions, which are much coarser grained than the resolution of our approach. For a typical program, such as each benchmark in SPEC CPU 2006, it needs at least $>$1000 and sometimes $>$10000 identifiers, which cannot be isolated by 256 regions, so TMDFI is not sufficient to support complete DFI for a typical program while our approach is.

There are also other tag-based isolation techniques. 
The work of \cite{Crandall04} uses 1-bit tag for each word of data to indicate its integrity 
level in Biba's low-water-mark integrity policy~\cite{Ken77}, which requires that 
an instruction can only modify data with integrity level no higher than that of the instruction.
In \cite{Crandall04}, processor hardware is modified to enforce this policy for
control data protection. 
In \cite{Watson15}, a 256-bit tag is employed to specify if each data can be referred by certain instructions. However, the 256-bit in \cite{Watson15} has a different meaning from the 16-bit identifier in our approach. For security, the approach in \cite{Watson15} only handles the permission of pointers. In contrast, our approach handles the permission of every store/load instruction.
Overall, the tag-based
techniques~\cite{Song16,Crandall04,Watson15,Liu18_2} provide only coarse-grained
isolation as different data/instructions with the same tag cannot be isolated from each other.

\begin{figure}[!hbt]
\begin{lstlisting}[basicstyle=\ttfamily\scriptsize, xleftmargin=2.5em,numbers=left]
struct vuln{
  char data[256];
  int off=0; 
  int size=0;
}*u0, *u1, *u2;
/* ============== */
char user0_buffer[256];                           //user0's buffer for storing u0's data
char user1_buffer[256];                           //user1's buffer for storing u1's data
char user2_buffer[256];                           //user2's buffer for storing u2's data
read_user_input(u0, user0_input);                 //store user0's input to u0
read_user_input(u1, user1_input);                 //store user1's input to u1
...
memcpy(user0_buffer+u0->off, u0->data, u0->size); //copy u0's data to user0's buffer
memcpy(user1_buffer+u1->off, u1->data, u1->size); //copy u1's data to user1's buffer
memcpy(user2_buffer+u2->off, u2->data, u2->size); //copy u1's data to user1's buffer
\end{lstlisting}
\vspace{-2ex}
    \caption{An example of vulnerability that HDFI cannot detect.}
    \label{fig:hdfi_ex}
\end{figure}

\section{Performance Overhead Analysis}
\label{sec:sys}

We analyze the source of performance overhead of software DFI~\cite{Castro06}. The experiment setup of the analysis is the same as that in Section{~\ref{sec:expsetup}}. We call the program to be checked by DFI enforcement the \textbf{user program}. 
For a user program, when each \texttt{store} or \texttt{load} is executed, RDT needs to be accessed and consequently data transfer with memory may be greatly increased. A memory access typically takes hundreds of clock cycles and can cause huge overhead. Thus, we first tested the cache hit rate to understand the DFI's impact on memory accesses.

\begin{figure*}[!hbt]
	\centering
	\includegraphics[width=1.0\textwidth]{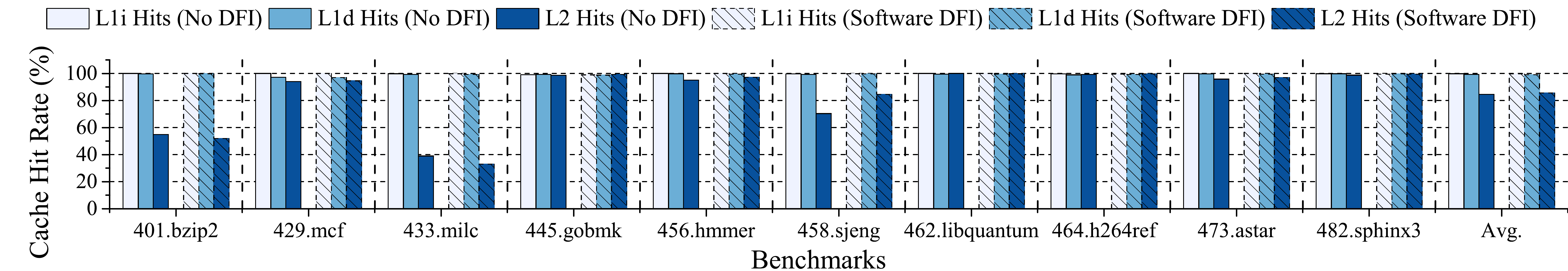}
	\caption{Cache hit rates of user programs with and without software DFI. The high cache hit rates indicate that memory access is probably not a bottleneck.}
	\label{fig:cache_hit}
\end{figure*}

When testing with SPEC CPU 2006 benchmark{~\cite{spec}}, the cache hit rates of user programs without DFI enforcement and with software DFI are shown in Fig.~\ref{fig:cache_hit}. One can see that the cache hit rates are usually greater than 95\% regardless of applying DFI enforcement or not. 
This indicates that memory access is probably not a bottleneck. 

 \begin{figure}[!hbt]
    \centering
    \includegraphics[width=0.65\textwidth]{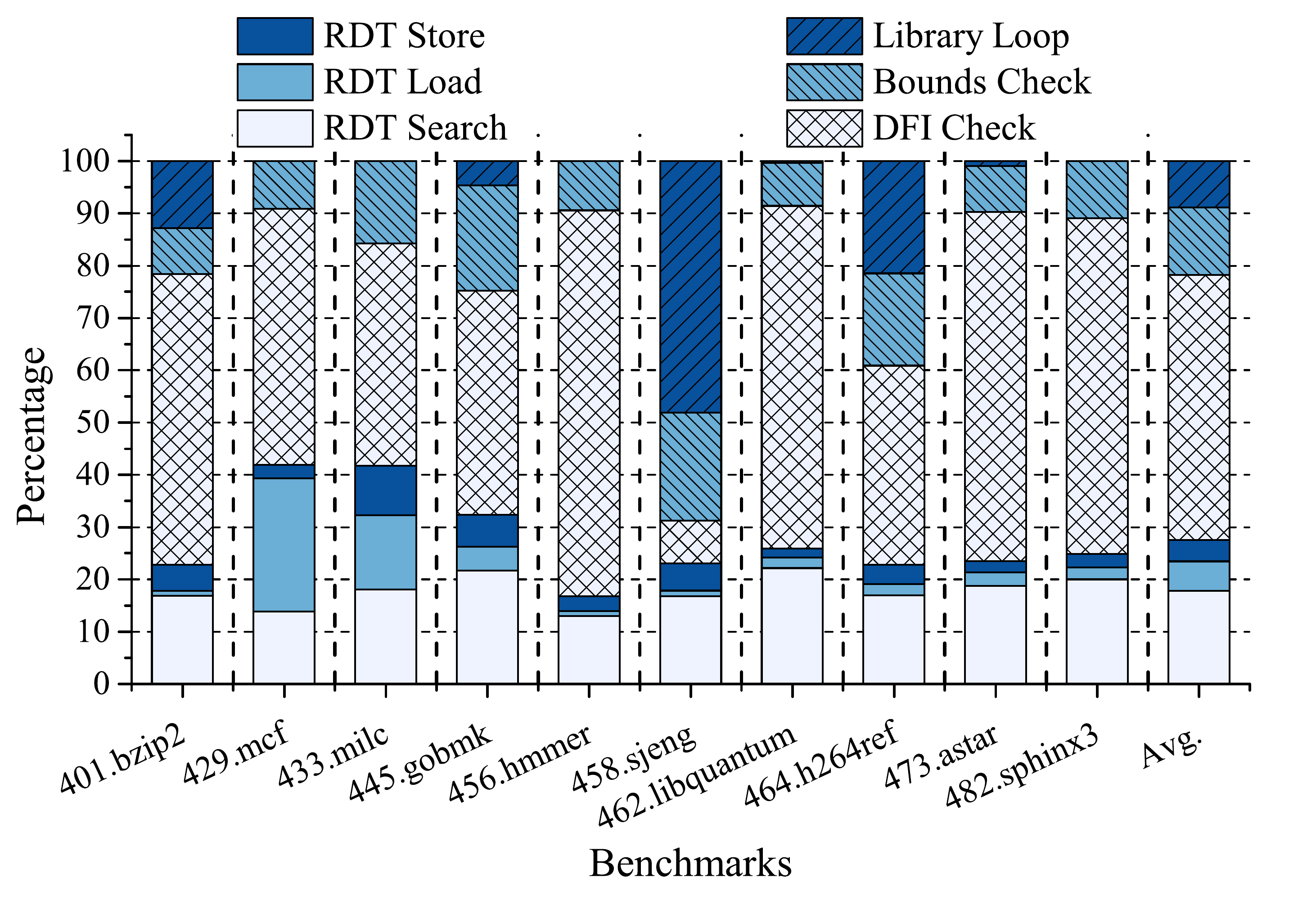}
    \caption{Overhead breakdown of software DFI. The results indicate that the bottleneck is not memory access but ``DFI check'' instructions.}
    \label{fig:ov_break}
    \vspace{-2ex}
\end{figure}

We further investigated the overhead breakdown of software DFI, of which the results are shown in Fig.~\ref{fig:ov_break}, where ``RDT Search'' represents the execution of the instrumented instructions for finding the RDT entry of the corresponding user \texttt{load} or \texttt{store}. ``Bounds Check'' means the check for preventing RDT from illegal modification. ``Library Loop'' represents the execution of the loop related instructions (such as comparison and branch instructions) in the instrumentation for each library function. ``DFI Check'' indicates the comparisons checking if the identifier found in RDT Search is in the RDS of the corresponding user \texttt{load} or not.

According to Fig.{~\ref{fig:ov_break}}, most of the overhead is from ``DFI check''. It also shows that RDT access (excluding ``RDT search'') contributes little to the overhead. 
This confirms that the bottleneck is not memory access but ``DFI check'' instructions. 
Specifically, many comparison and branch instructions are executed for each ``DFI check'', which compares the identifier found in RDT with each identifier in RDS of the corresponding user \texttt{load}. 
Although this check computation is fairly simple, it is performed for a huge volume of data.

\section{Overview of Proposed Approach}
\label{sec:overview}

Our approach is to delegate DFI enforcement to
another computing resource external to the main processor where the user program is executed. The delegated resource 
can be a processor core in a Chip Multiprocessor (CMP) or
a Processing-In-Memory (PIM) processor~\cite{pimgem5paper}. The two options are similar in terms of the overhead reduction. 
We will use PIM as an example platform to describe our approach while the same idea is applicable to the CMP core option.

The PIM processor undertakes most of the DFI verification components analyzed in Section{~\ref{sec:background}}, and
can quickly access RDSs and RDT in its vicinity.
As such, what remains for
the main processor to do is to collect target instruction information discussed in Section~\ref{sec:background} and send it to PIM. 
Although the information collection and transmission can be implemented with software in a way same as multithreading,
our study shows that such a software approach still
experiences huge or even worse performance overhead.
Thus, we propose a hardware approach to minimize 
extra software executions at the main processor.

\begin{figure}[!hbt]
	\centering
	\includegraphics[width=0.6\textwidth]{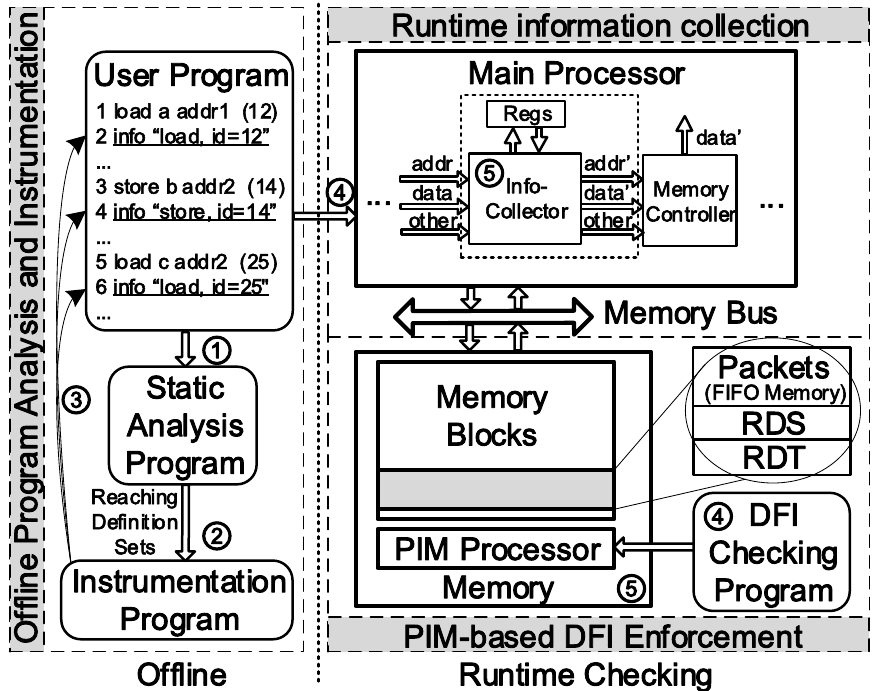}
	\caption{The proposed system and the flow of PIM-based DFI enforcement.}
	\label{fig:flow}
\end{figure}

The proposed system is depicted in Fig.{~\ref{fig:flow}}, which consists of 3 main blocks:
\begin{enumerate}[itemsep=0ex]
    \item Offline program analysis and instrumentation (Section~\ref{sec:instru}).
    \item Runtime information collection (Section~\ref{sec:hardware}).
    \item PIM-based DFI Enforcement (Section~\ref{sec:checkprog}).
\end{enumerate}

\textbf{Offline Program Analysis and Instrumentation:} According to Section{~\ref{sec:background}}, RDSs of the user program are required by DFI enforcement. Since RDS of each instruction is static, offline program analysis can be applied to the user program once and RDSs can be loaded into the memory when the user program starts. Besides, the software instrumentation is introduced for generating the target instruction information, which can be used by the hardware in the Runtime Information Collection block. In Fig.{~\ref{fig:flow}}, the underlined instructions are for instrumentation. One example is \texttt{info "load, id=12"}, which indicates the former instruction is \texttt{load}, with the identifier 12.

\textbf{Runtime Information Collection:} The target instruction information generated from the instrumentation needs to be transferred to the PIM processor. The information transfer is performed by the dedicated hardware module named \textbf{info-collector} designed in the main processor. Info-collector parses the instrumented instruction for the target instruction information, and it can optimize the size of the information, which is sent to the memory while the software program is being executed.

\textbf{PIM-based DFI Enforcement:} This block contains the PIM processor, which performs the DFI checking after receiving the target instruction information from the main processor and accessing RDS and RDT in the memory. With the collaboration of the three blocks, complete runtime DFI enforcement is realized.

\textbf{Putting It All Together:} Fig.{~\ref{fig:flow}} shows the overall flow of the proposed DFI enforcement, where the circled numbers indicate the step ID:
\begin{enumerate}[itemsep=0ex,label=\protect\circled{\arabic*}]
\item Static analysis is performed for a user program.
\item RDSs are obtained from the static analysis.
\item The codes are instrumented automatically. The main instrumentation is to add
instructions for encoding the target instruction information after each target instruction so as to help collect its information. The instrumented instructions are underscored in Fig.~\ref{fig:flow}.
\item The DFI checking program and RDS are loaded onto the PIM processor before the user program execution starts on the main processor. 
\item During program execution, the info-collector parses each instrumented instruction, collects target instruction information accordingly, forms a
\textbf{DFI packet}, and sends it to the PIM processor, where enforcement computations are performed or RDT is updated.
\end{enumerate}

In the following sections, the details of the three blocks are elaborated.

\section{Software Instrumentation}
\label{sec:instru}

Instrumentation is to add additional code into a user program in order to facilitate
the DFI enforcement. 
The software instrumentation in our approach
helps not only extract the necessary information but also avoid changing the instruction set. 

As shown in the Offline Program Analysis and Instrumentation block in Fig.{~\ref{fig:flow}}, a user program is automatically instrumented by the software we developed after getting the reaching definition sets (RDSs) from the static analysis. The instrumentation mainly adds the instructions that generate the runtime information of target instructions, which can be parsed by the info-collector. Assuming the instrumented instruction is \texttt{info}, the basic syntax is
\begin{center} \vspace*{-1mm}
\small\texttt{info    runtime\_info} \vspace*{-1mm}
\end{center}
When \texttt{info} is executed in the main processor, the \texttt{runtime\_info} is transferred to the info-collector. A simple way to realize the \texttt{info} instruction is to extend the instruction set. However, instruction set architecture (ISA) extension requires much more changes across the computing stack including both software and hardware, thereby introducing more engineering efforts and cost.
To avoid this, we propose another approach to implement the \texttt{info} instruction, by overloading the \texttt{store} instruction.
These instrumentation \texttt{store}
instructions are called \textbf{DFI \texttt{store}}, 
of which we overload the use with underlying semantics different from ordinary \texttt{store} instructions. 
Our key technique is to
differentiate between ordinary \texttt{store}
and DFI \texttt{store} without adding new instructions. 
The basic syntax of the DFI \texttt{store} is
\begin{center} \vspace*{-1mm}
\small\texttt{store    runtime\_info  dfi\_global} \vspace*{-1mm}
\end{center}
where \texttt{runtime\_info} is a constant value including the runtime information, and \texttt{dfi\_global} is the address (the pointer) of a global variable
declared at the beginning of a program and
serves as a signature to indicate a
DFI \texttt{store}. The address of this
global variable is set by writing a dummy value
at the beginning of a program as
\begin{center} \vspace*{-1mm}
\small\texttt{store  dfi\_dummy dfi\_global} \vspace*{-1mm}
\end{center}
The \texttt{dfi\_dummy} is a dummy value that has a fixed value 
to obtain the destination address of \texttt{dfi\_global}. The info-collector can obtain \texttt{dfi\_global} by identifying the first \texttt{store} in a program that stores \texttt{dfi\_dummy} to an address. For example, \texttt{dfi\_dummy} can be designed as \texttt{123456}. Once \texttt{store 123456 11122} is executed, the info-collector assigns \texttt{11122} to \texttt{dfi\_global}, and \texttt{dfi\_global} can only be assigned one time for a user program.

The info-collector (dotted box in Fig.{~\ref{fig:flow}}) checks if the target address of a \texttt{store} instruction 
is the same as \texttt{dfi\_global}.
If so, the instruction is a DFI \texttt{store} and the runtime information is extracted and sent to PIM.

There are three scenarios where the instrumentation is needed:
\begin{enumerate}[itemsep=0ex]
    \item For each ordinary \texttt{store} or \texttt{load} instruction, its target instruction information is required by DFI, thereby instrumentation is needed.
    \item The source code of a library function is not necessarily accessible to the users, but instrumentation can still be performed to obtain the target instruction information, if the library functions are for memory accesses. This is similar to the wrapper{~\cite{Castro06}} in theory, but our implementation is hardware-based but not software-based.
    \item Function return addresses are stored on stack and vulnerable to attacks such as Return-Oriented Programming (ROP){~\cite{rop}}. We treat their accesses as implicit \texttt{load/store} instructions and perform DFI enforcement accordingly. When a parent function \texttt{parent\_func()} calls a child function \texttt{child\_func()}, the return address is stored on the stack by an instruction \texttt{parent\_inst}. When function \texttt{child\_func()} returns, the return address is loaded by a return instruction \texttt{child\_inst}. DFI ensures that the return address used by \texttt{child\_inst} should be the latest value stored by \texttt{parent\_inst}. To enforce the DFI of function return addresses, we need instrumentation for generating the target instruction information of function calls and returns.
    
\end{enumerate}

\subsection{Instrumentation for DFI Enforcement}
\label{sec:instru_pro}

The instrumentation is mainly to extract 
the runtime information of the \texttt{load}/\texttt{store} instructions
in a user program related to DFI checking,
and sent to the PIM processor.
The information includes instruction identifier, instruction type and
target address of \texttt{load}/\texttt{store}. Instruction identifiers
are automatically assigned by the instrumentation tool. 

Every \texttt{store} and \texttt{load} instruction in a user program, called target instruction, is followed by
a DFI \texttt{store}. The \texttt{runtime\_info} of the DFI \texttt{store} contains
the instruction type and identifier of the preceding target 
instruction. For example, in Fig.~\ref{fig:flow}, line 2 is an instrumentation
instruction \texttt{info ``load, id=12''} that is implemented by \texttt{store ``load, id=12'' dfi\_global}, which tells 
the instruction type and identifier of the target instruction 
in line 1. To encode the instruction type and identifier, according to~\cite{Castro06}, 16 bits are sufficient for representing instruction identifiers in a large program.
We use an additional bit to indicate instruction type, where 0 means \texttt{write} and 1 means \texttt{read}. 
When the info-collector recognizes a
DFI \texttt{store}, it extracts the target address of
the preceding target instruction. The target address and 
the \texttt{runtime\_info} form a {\bf DFI packet} to be sent 
to PIM. 

At the beginning of code execution,
a memory space is dynamically allocated at the PIM processor for DFI
enforcement. This includes the memory space for storing incoming packets, which is called packet \textbf{FIFO memory}.
The starting address of packet FIFO memory is \texttt{packet\_mem\_addr}, which is also a dynamic value. 
Similar to \texttt{dfi\_global}, \texttt{packet\_mem\_addr} is also set by writing a dummy packet at the beginning of a program as
\begin{center} \vspace*{-1mm}
\small\texttt{store packet\_dummy packet\_mem\_addr} \vspace*{-1mm} 
\end{center}
\begin{sloppypar}

Later during the code execution, all DFI packets are
sent to FIFO memory based on \texttt{packet\_mem\_addr}. Please note that
\texttt{dfi\_global} and \texttt{packet\_mem\_addr} are generated
by the automatic code instrumentation, and not visible to
security attackers. Besides, only the info-collector is allowed to control the memory controller to send data to the FIFO memory, after \texttt{packet\_mem\_addr} is set. Any other attempts for accessing the FIFO memory in the main processor are identified as violations.
\end{sloppypar}

\begin{figure}[!hbt]
\begin{lstlisting}[basicstyle=\ttfamily\scriptsize, xleftmargin=3em,numbers=left]
/* =====beginning of the program====== */
<@\mysize\underbar{(instructions for allocating FIFO memory)}@>
<@\mysize\underbar{(instructions for storing RDS to memory)}@>
<@\mysize\underbar{\textbf{\color{blue}store} dfi\_dummy dfi\_global}@>
<@\mysize\underbar{\textbf{\color{blue}store} packet\_dummy packet\_mem\_addr}@>
...
store x1 addr1       //identifier: 12
<@\mysize\underbar{\textbf{\color{blue}store} (0$<<$16)+12 dfi\_global}@>
...
load x2 addr2        //identifier: 25
<@\mysize\underbar{\textbf{\color{blue}store} (1$<<$16)+25 dfi\_global}@>
\end{lstlisting}
    \vspace{-1em}
    \caption{An example of code instrumentation.}
    \label{fig:instru}
    \vspace*{-1mm}
\end{figure}

An example of the instrumentation is shown in Fig.~\ref{fig:instru}
where lines 7 and 10 are the original instructions in the user program, while lines 2, 3, 4, 5, 8 and 11 are instrumentations. The identifiers of the instructions at lines 7 and 10 are in the parentheses (12 and 25). 
The data of a DFI \texttt{store} (lines 8 and 11 in Fig.~\ref{fig:instru}) has bit 16 for instruction type and bits 15-0 for an instruction identifier.

\subsection{Handling Library Functions}
\label{sec:lib}

A program often calls library functions, whose source code is not necessarily directly accessible. 
This makes it hard to directly instrument the DFI \texttt{store}s inside the library functions. 
The seminal work{~\cite{Castro06}} proposed a wrapper approach for solving this problem, and we propose its implementation scheme to effectively enforcing DFI for library functions, by instrumenting the DFI \texttt{store}s outside a library function.
As a library function call
may involve a multi-byte data block in general, the instrumentation
needs to keep track of data-length besides data address. To include these kinds of information in \texttt{runtime\_info} of the DFI \texttt{store}, multiple DFI \texttt{store}s are required.
In detail, our approach is illustrated using the example in 
Fig.~\ref{fig:instru_lib}.

\begin{figure}[!hbt]
\begin{lstlisting}[basicstyle=\ttfamily\scriptsize, xleftmargin=3em,numbers=left]
<@\mysize\underbar{\textbf{\color{blue}store} (1$<<$20)+(1$<<$19)+(0$<<$18)+(1$<<$17)+7 dfi\_global}@>
<@\mysize\underbar{\textbf{\color{blue}store} (y1's addr) dfi\_global}@>
<@\mysize\underbar{\textbf{\color{blue}store} (x1's addr) dfi\_global}@>
<@\mysize\underbar{\textbf{\color{blue}store} 40 dfi\_global}@>
memcpy(x1, y1, 40)           //identifier: 7
...
<@\mysize\underbar{\textbf{\color{blue}store} (1$<<$20)+(0$<<$19)+(1$<<$18)+(1$<<$17)+15 dfi\_global}@>
<@\mysize\underbar{\textbf{\color{blue}store} (x2's addr) dfi\_global}@>
<@\mysize\underbar{\textbf{\color{blue}store} 12 dfi\_global}@>
<@\mysize\underbar{\textbf{\color{blue}store} 9 dfi\_global}@>
memset(x2, 3, (9<<32)+12)    //identifier: 15
\end{lstlisting}
    \vspace{-1em}
    \caption{The instrumentation for library functions.}
    \label{fig:instru_lib}
    \vspace*{-1mm}
\end{figure} 

In this example, the target instructions are the function calls
in lines 5 and 11, with their identifiers in parentheses. The instrumentation for each library
function call includes multiple DFI \texttt{store} instructions
like lines 1-4 for the target instruction of line 5. 
The first DFI \texttt{store} keeps the corresponding
identifier in its lower 16 bits. 
Its bits 17-20 are four binary indicators telling if the target instruction is a library function call or not, if the data-length needs 64 bits to represent or not, and if the function loads/stores data or not.
The info-collector parses these indicators and then takes 
corresponding actions. Additional DFI
\texttt{store} instructions are added to send other information. For example, lines 2 and 3 send load
and store addresses. Depending on if the data-length is represented
in 32 or 64 bits, the data-length needs to be sent through 
a single or two DFI \texttt{store} instructions.
For example, line 4 sends the data-length in a single
DFI \texttt{store} while lines 9 and 10 send in
two DFI \texttt{store} instructions. 

If the arguments of a library function call do not include the data address or data-length, or the function is called by an indirect branch instruction, the approach in Fig.{~\ref{fig:instru_lib}} cannot be directly applied. Instead, more complex instrumentation schemes and library function overriding are needed. Since enforcing DFI for library functions is not the main focus of this work, we leave it as the future work.

\section{Hardware Design}
\label{sec:hardware}

In Fig{~\ref{fig:flow}}, the info-collector in the Runtime Information Collection block is the key hardware component to be added to the main processor. It detects DFI \texttt{store} instructions, collects runtime information of target instructions, generates DFI packets and sends them to PIM. Due to the existence of three instrumentation scenarios, the info-collector needs to first identify the scenario and generate a DFI packet accordingly. 
The DFI packet is then sent to the FIFO memory.
Besides, since transferring data to the memory can lead to performance overhead, data compression and runtime optimization are applied.
The design details are described in the following subsections.

\subsection{DFI Packet Generation}
\label{sec:dfi_op}

\begin{figure}[!hbt]
	\centering
	\includegraphics[width=0.6\textwidth]{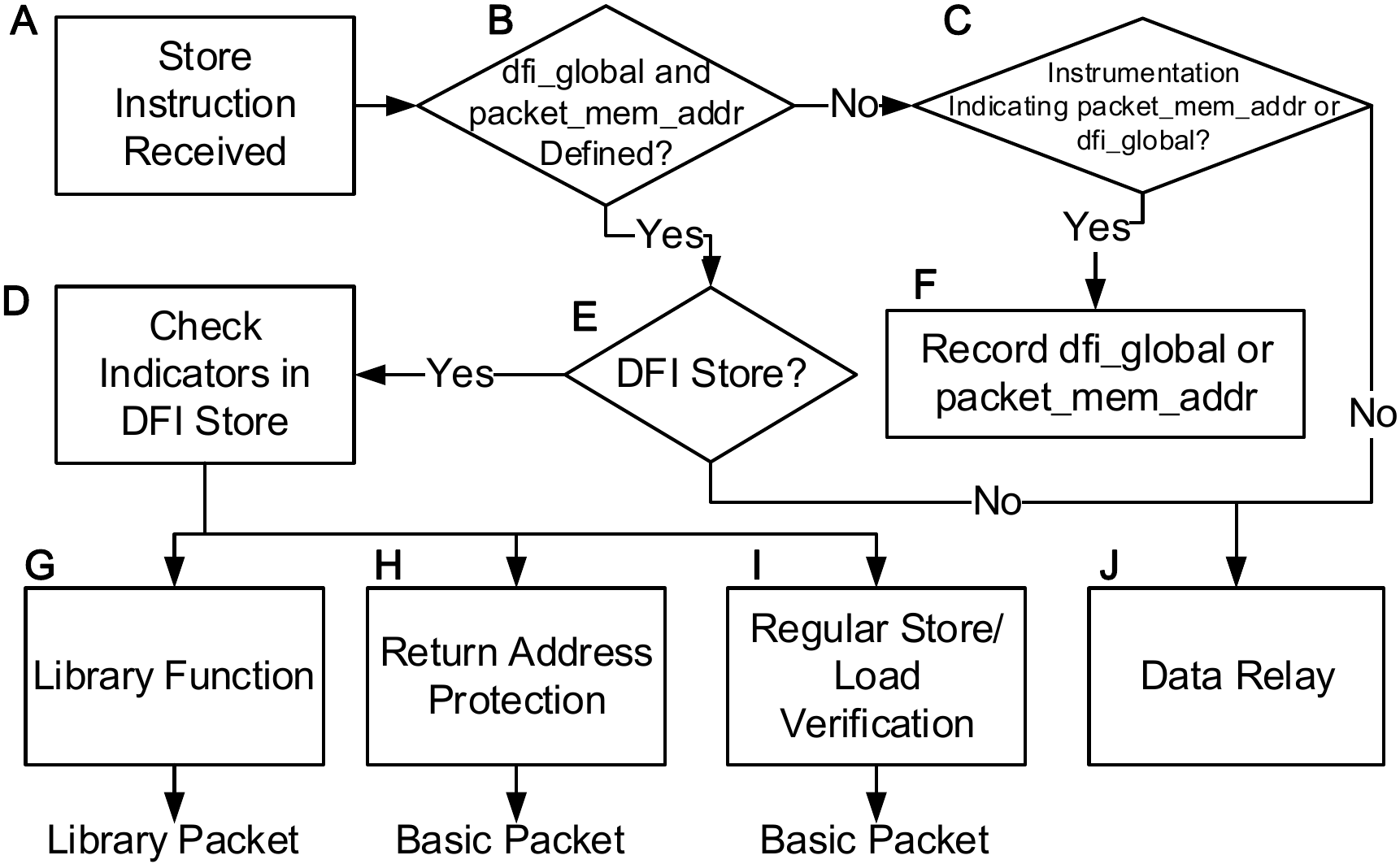}
	\caption{Operations of info-collector.}
	\label{fig:dfi_op}
\end{figure}

Info-collector can be realized as a hardware
circuit through synthesizing Verilog description. 
Its basic operations are depicted in Fig.~\ref{fig:dfi_op}.

\begin{sloppypar}
The info-collector acts only when a \texttt{store} instruction is
executed. In step B of Fig.~\ref{fig:dfi_op}, it checks if 
\texttt{dfi\_global} and \texttt{packet\_mem\_addr} have already been defined.
If not, it proceeds to step C to capture \texttt{dfi\_global} or \texttt{packet\_mem\_addr}.
Please note ``\texttt{store dfi\_dummy dfi\_global}'' and
``\texttt{store packet\_dummy packet\_mem\_addr}'' are instrumented
at the beginning of a program. 
Moreover, both \texttt{dfi\_dummy} and \texttt{packet\_dummy} have
signature values that can be recognized by the info-collector. 
If they have already been defined, the info-collector further
checks if the \texttt{store} is a DFI \texttt{store}.
This is by examining if the target address is the same as that of \texttt{dfi\_global}.
\end{sloppypar}

If this \texttt{store} is a DFI \texttt{store}, the info-collector 
parses the indicators in the data part of the DFI \texttt{store} and tells if this is to verify 
\texttt{load/store} for DFI enforcement, function return or a library function call. 
If this instrumentation is for a \texttt{load/store} instruction, the info-collector 
collects instruction type and identifier from this DFI \texttt{store} instruction,
and the target address from the previous instruction.
These pieces of information form a {\bf basic packet} (\texttt{data'} in Fig.~\ref{fig:flow}) to be sent to PIM, which stores the packet to the address of the allocated packet FIFO memory (\texttt{addr'} in Fig.~\ref{fig:flow}).

If this DFI \texttt{store} is for a return address protection (step H in Fig.~\ref{fig:dfi_op}), the info-collector takes the identifier and instruction type from this DFI \texttt{store}, 
and extracts the pointer to the return address from the next DFI \texttt{store}.  This information also forms a \textbf{basic packet}.
If this DFI \texttt{store} is for a library function (step G), the indicators of this \texttt{store} tell if the library function is to load data, store data or not, and if the data-length needs to be encoded by 64 bits or not. Next, the info-collector continues to collect additional information from subsequent DFI \texttt{store} instructions and generates a \textbf{library packet} to be sent to PIM.

If the \texttt{store} instruction is a part of the user program (step J), i.e., not a
DFI \texttt{store}, its \texttt{data} is relayed to memory without any change and
its target address is stored in a local register for future use.

\subsection{Packet Transfer to PIM}
\label{sec:transfer}

A memory space is allocated to store DFI packets sent from the main processor. 
It is used as a packet FIFO to store and process the packets in a first-come-first-serve
manner. In order to maintain the FIFO nature using a region of random access memory with
low overhead, we develop circuit design techniques to maintain the head and tail pointers
in hardware, where the head pointer is updated by PIM (consumer) and tail pointer is
updated by the main processor (producer).

\subsection{Lossless Data Compression}
\label{sec:dfi_cp}

The main reason for performance overhead of PIM-based DFI is transferring DFI packets to memory. 
 Although each DFI packet has only a few bytes, the number of DFI packets is huge and the overall impact is significant.
We propose to compress
target addresses and identifiers by exploiting locality. The compression is realized in the 
info-collector hardware.

Consider the two C program examples in Fig.~\ref{fig:dfi_cpex}. 
For example A, assume the starting memory address of \texttt{aa} is 0x8000, then the program stores data at 0x8000, 0x8004, 0x8008, and so on. Starting from \texttt{i=1}, each target address 
increases by 4 compared to the previous one. Thus, we only need to send the increment in 4 bits,
which include 1 sign bit, instead of a 32-bit address.
Example B in Fig.~\ref{fig:dfi_cpex} is similar, but has an address pattern of 
0x8000, 0x8400, 0x8800, etc. Although the address
increment 0x400 is relatively large and needs 11 bits to represent, the lower bits of the increment
are all 0s. Thus, instead of using integer compression, we use a format similar to floating point number representation to further reduce the bitwidth of the 
address increment. This format consists of a sign bit, significand and exponent of 16. 
To represent 0x400, the sign bit is 0, there are 3 bits for significand to represent 4 and 
the exponent is 2. Overall, the bitwidth is 6, which is shorter than the 11-bit binary encoding. 
The floating point number representation contains 8-bits, 1 sign bit,
4 bits of significand and 3 bits of exponents (the power of 16). This representation can cover the
range from $-15\times 2^{28}$ to $15\times 2^{28}$. The info-collector calculates
the difference between two target addresses. If the difference is within this range and the significand is within $-15$ to $15$, then the difference is represented by an 8-bit floating point number. Note that the difference is compressed only when it can be represented in this format with a 16-basis exponent.

Identifiers can also be compressed based on their value locality. However, 
they rarely have the patterns like example B, where the increment is at the middle bits
of an address. Thus, the difference between two identifiers is represented by a binary number.
Overall, a DFI packet can be compressed to 15 bits. Thus, we can pack two 
\textbf{compressed packet}s into one word. 

\begin{figure}[!hbt]
\begin{lstlisting}[basicstyle=\ttfamily\scriptsize, xleftmargin=2.5em,numbers=left]
/* ========Example A========== */
int aa[1024];
for(int i=0;i<1024;i++)
  aa[i]=i;
/* ========Example B========== */
int bb[1024][1024];
for(int i=0;i<1024;i++)
  for(int j=0;j<1024;j++)
    bb[j][i]=i+j;
\end{lstlisting}
    \vspace{-1em}
    \caption{Examples of address locality.}
    \label{fig:dfi_cpex}
    \vspace*{-1mm}
\end{figure}

\subsection{Runtime Optimization}
\label{sec:dfi_buf}

We develop packet pruning techniques and a technique for increasing the opportunity of locality for 
data compression. These optimization techniques help reduce the amount of data 
sent to PIM and thereby further decrease performance overhead. 
Some pruning techniques described here are similar to those in \cite{Castro06}. 
However, the pruning techniques in \cite{Castro06} are offline while our hardware
approach allows pruning at runtime. As more information, such as target address,
is available at runtime, the opportunity of pruning is increased.

Similar to data transfer between memory and cache in cache lines, we pack multiple DFI packets into 
a block of hundreds of bytes before sending them to PIM. The packets in a block
are organized in a \textbf{transmission buffer}, which is implemented as a register file. 
The optimizations are performed for packets in the buffer before they are
sent out. Note that waiting for other packets to form a block increases
DFI enforcement latency but does not increase performance overhead.

\begin{figure}[!hbt]
	\centering
	\includegraphics[width=0.9\textwidth]{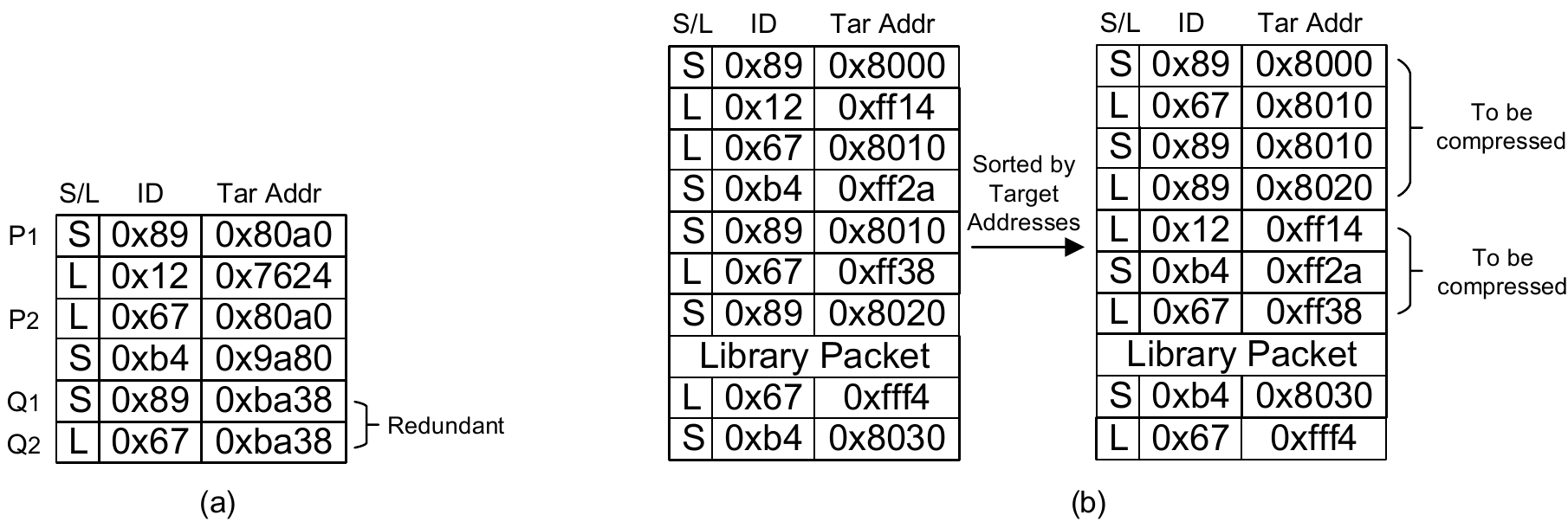}
	\caption{(a) The example for illustrating optimization D.  (b) The example for illustrating optimization E.}
	\label{fig:optdeex}
\vspace{-2ex}
\end{figure}

Consider two pairs of basic packets in the transmission buffer, $(P_1, P_2)$
and  $(Q_1, Q_2)$.  Each basic packet is for instruction \texttt{load},
 \texttt{store}, or function return.
Packet $P_1$ ($Q_1$) precedes $P_2$ ($Q_2$).
The packets of each pair share the same target address and there is 
no other DFI packet for \texttt{store} of the same target address between them. 
There are five optimization techniques described using the packet pairs:

\begin{enumerate}[itemsep=0ex,leftmargin=1.8em, label=\Alph*:]
\item If $P_1$ and $P_2$ are for \texttt{store} instruction, and there is no 
other DFI packet for a \texttt{load} with the same target address between them, then 
packet $P_1$ is redundant and can be pruned out without being sent to PIM.
\item If $P_1$  and $P_2$ are both for \texttt{store} instruction, and their identifiers are the same, 
then $P_2$ can be pruned out.
\item If $P_1$ and $P_2$ are both for \texttt{load} instruction, and their identifiers are the same, 
then $P_2$ can be pruned out.
\item $P_1$/$P_2$ are for \texttt{store}/\texttt{load} of the same target address $Addr_1$. 
After $P_1$ and $P_2$, packets $Q_1$ and $Q_2$ are for \texttt{store}/\texttt{load} of the same target address $Addr_2$. $P_1$/$P_2$ have identifiers $\alpha$/$\beta$, respectively. If $Q_1$/$Q_2$ also have identifiers $\alpha$/$\beta$, respectively, then $Q_1$ and $Q_2$ are redundant. This is to make sure that the same \texttt{store/load} pair appears only once in the transmission buffer. An example is shown in Fig.{~\ref{fig:optdeex}}(a), where the table is the packets in the transmission buffer, with each line representing a basic packet. The last line represents the latest packet. ``S/L'' represents the instruction type (``S'' for \texttt{store} and ``L'' for \texttt{load}), and ``Tar Addr'' represents the target address. In this example, $Q_1$ and $Q_2$ are redundant.
\item All basic packets in the transmission buffer are sorted according to their target addresses.
If two packets have the same target address, their relative order keeps unchanged. 
If there is a library packet, the basic packets before and after this library packet are sorted separately. 
After sorting, the target address difference between two adjacent packets is examined to 
find if data compression can be performed. The sorting helps find opportunities for 
data compression. The verifications in DFI enforcement for \texttt{load/store} of different target addresses 
are independent of each other and hence sorting does not affect DFI enforcement results. An example is shown in Fig.{~\ref{fig:optdeex}}(b), where the left table is the packets in the transmission buffer before sorting. Before sorting, the difference of the target addresses between each pair of adjacent packets is large, which is hard for compression. After sorting, we found two groups of packets that can be easily compressed by our compression approach. Note that the sorting is performed before and after the library packet separately.
\end{enumerate}

Among the optimizations, A, B and C are similar to those in \cite{Castro06}
except that they can be performed both offline and at runtime while those in \cite{Castro06} are restricted to offline.
Techniques D and E are newly developed in this work. After the optimizations are performed,
a packet is compressed if possible.

\subsection{Circuit Implementation of the Optimizations}
\label{sec:hardimpl}

All the 5 optimizations can be realized in circuits for runtime
use in the main processor. We illustrate the circuit designs
by using optimization C as an example.

\begin{figure}[!hbt]
	\centering
	\includegraphics[width=0.5\textwidth]{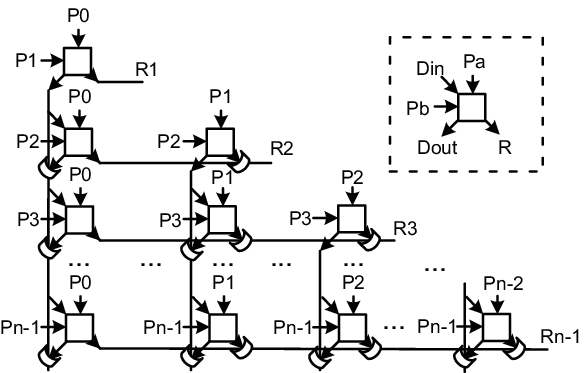}
	\caption{Circuit for implementing optimization C.}
	\label{fig:opt3}
\vspace{-2ex}
\end{figure}

The schematic of combinational circuit implementation of optimization C is shown in Fig.~\ref{fig:opt3}. Assume there are $n$ basic packets in the transmission buffer, $Pi$ represents the $i$-th packet, 
and $Ri$ indicates if the $i$-th packet is redundant or not. 
Each square in Fig.~\ref{fig:opt3} is a Processing Element (PE) that computes if a packet is redundant or not. 
In each column of Fig.~\ref{fig:opt3}, a packet 
$Pi$ is compared with all later packets $Pj, j > i$ and attempts
to find a redundant $Pj$ to be pruned. If there are multiple packets that
are redundant with respect to $Pi$, only the topmost one
(with the smallest $|j-i|$) is asserted for pruning and the others 
can be pruned later in other columns to the right. 
The $R$ signals in a row are $OR$ed such that a packet in a row 
can potentially be pruned by any proceeding packets organized in columns. 
For example, $P3$ in row 3 can be potentially pruned by 
$P0$, $P1$ or $P2$ in the left three columns. 
Like illustrated in the dotted box, 
a PE compares two input packets $Pa$ and $Pb$.
A necessary but insufficient condition for asserting $R=TRUE$ is that
$Pa$ and $Pb$ are both for \texttt{load} with the same 
target address and identifier. The final result of $R$ also 
depends on $Din$, which is a disable signal for the pruning.
The value of $R=TRUE$ when $Din == 0$ and the necessary condition holds.
There are two scenarios where the disable signal asserts:
(1) there is a \texttt{store} at the same target address between
the two \texttt{load} instructions of $Pa$ and $Pb$, and thus
the conditions for optimization C is not completely satisfied;
(2) a redundant packet has already been found and no further pruning
is needed in a column. 
For scenario (1), $Dout = 1$ when $Pa$ is for \texttt{load} while
$Pb$ is for \texttt{store}. 
For scenario (2), $Dout = 1$ if $R=TRUE$ for the same PE.

\section{DFI Checking Program at PIM}
\label{sec:checkprog}

In the PIM-based DFI Enforcement block in Fig.{~\ref{fig:flow}}, the DFI checking program at the PIM processor continuously reads DFI packets from the FIFO memory, and either performs DFI enforcement or updates RDT. For different types of DFI packets, the PIM processors take different actions.

The DFI checking program is written in C language, and its binary code 
is executed on the PIM processor. 
The RDT memory space is allocated by the instrumentation code.
Same as in \cite{Castro06}, all program data are organized in words, each of which requires one RDT entry. 
 If the data memory for the user program has $N$ bytes, there are $N$/$4$ entries in the RDT~\cite{Castro06}. 
 Since each identifier has 16 bits = 2 bytes, the RDT uses $\frac{N\times2}{4}=N$/$2$ bytes of memory.

There are three kinds of DFI packets to be processed by the DFI checking program.
\begin{itemize}[itemsep=0ex,leftmargin=1em]
\item \textbf{Basic packet for \texttt{store} or \texttt{load}:} 
The DFI checking program extracts instruction type, identifier $\alpha$ and target address $\beta$ from the packet. 
If the instruction type is \texttt{store}, identifier $\alpha$ is stored at entry $\beta>>2$ of RDT. 
The right shift is performed because RDT is organized in words. 
If the instruction type is \texttt{load}, the DFI checking program reads 
identifier $\gamma$ from entry $\beta>>2$ of RDT, and loads the RDS of identifier $\alpha$. 
Then, the program checks if $\gamma$ is in the RDS of $\alpha$ or not. If not, a DFI violation is reported. 
Finally, identifier $\alpha$ and target address $\beta$ are saved in registers for future decompression of
compressed packets.
\item \textbf{Compressed packet for \texttt{store} or \texttt{load}:} 
The process is similar to handling basic packets except that decompression is performed.
\item \begin{sloppypar} \textbf{Library packet:} 
The DFI checking program extracts target address $\alpha$ if there is \texttt{load} in the library function call, 
and target address $\beta$ if there is \texttt{store}. Then, data-length $\gamma$ (in words) of the \texttt{load} and/or \texttt{store} 
and identifier $\delta$ of this function are also extracted. If there is an address $\alpha$, 
the DFI checking program loads the identifiers $\epsilon_0$, $\epsilon_1$... $\epsilon_{\gamma-1}$ from entries $\alpha>>2$, $(\alpha>>2)+1$, ... $(\alpha>>2)+\gamma-1$ in 
the RDT, and checks if every $\epsilon_i$ is in the RDS of identifier $\delta$. 
If there is address $\beta$, the program stores identifier $\delta$ to all the entries from $\beta>>2$ to $(\beta>>2)+\gamma-1$ in the RDT.
\end{sloppypar}
\end{itemize}

\section{Experiment}
\label{sec:exp}
\subsection{Experiment Setup}
\label{sec:expsetup}

This section describes the experimental setup and the modeling and evaluation methodology.

\textbf{Software Analysis and Instrumentation:} The programs used in our paper are based on C/C++ and compiled by LLVM{~\cite{llvm}} and the static analysis is performed by an extended SVF{~\cite{sui2016svf}}. The instrumentation is performed on the program's LLVM Intermediate Representation (IR) by our software without interacting with LLVM. Then, the instrumented program is further compiled into binary code. 
Besides, our techniques are general and directly applicable to other programming languages supported by LLVM. Note that compilers, static analysis tools, and the static analysis itself is out of the scope of this work.

\textbf{System Configuration and Modeling:} We evaluate our approach and the proposed techniques using architecture simulations
through SMCsim~\cite{pimgem5,pimgem5paper}, which is an extension to the gem5
simulator~\cite{gem5} for accommodating PIM.
The main processor is an ARM Cortex-A15 with 2GHz frequency,
32KB L1 instruction cache, 64KB L1 data cache, 2MB L2 cache, and 512 MB memory. 
A single PIM processor is used and operates at 2GHz frequency~\cite{Yang19,Pugsley14}. 64MB memory is allocated for RDT, which is sufficient for the testcases in our experiment.
Other details of the PIM can be 
found in \cite{pimgem5paper,pimgem5}. Please note that the PIM configuration has little 
impact on the user program execution.

\textbf{Security and Performance Evaluation:} For security analysis, we used the RIPE benchmark suite~{\cite{ripepaper}} for control-data attacks, and tested on Heartbleed vulnerability~{\cite{heartbleed}} and Nullhttpd heap overflow vulnerability~{\cite{nullhttpd}} for non-control data attacks.
In addition, we used SPEC CPU 2006~{\cite{spec}} benchmark suite for performance evaluation.

\subsection{Security Analysis}
\label{sec:se}

Our approach enforces the same DFI as defined in \cite{Castro06} and thus 
achieves similar security as \cite{Castro06} except that 
our approach is asynchronous monitoring~\cite{XiaDSN12,DasTIFS16,Lee17}, where detection
of DFI violation can trigger system interrupt
for further security measures,
rather than synchronous enforcement like \cite{Castro06}. 
This difference is a tradeoff between security and service availability.
Synchronization inevitably entails extra performance overhead as DFI enforcement blocks user program executions.

\subsubsection{RIPE Benchmark} 

RIPE~\cite{ripegit, ripepaper} is a well-known benchmark containing various control-flow attacks, and all control-flow attacks can be identified by DFI. RIPE is originally designed for X86 architecture and modification is required for executions on an ARM processor. 
We implemented 156 attacks of the benchmark for our system, including Return-Oriented Programming (ROP) \cite{rop} attacks and Jump-Oriented Programming (JOP) \cite{jop} attacks.
Table{~\ref{tab:ripeconfig}} shows the configuration dimensions and possible configurations of RIPE, where "Overflow Technique" indicates whether the attack target can be directly reached by sequentially overflowing from a buffer. "Attack Code" is what the attack payload is. "Target Code Pointer" is the target to be attacked. "Location" is the location of the attack target. ``Function Abused'' is the function used to modify the data{~\cite{ripepaper}}.

\begin{table}[!htb]
\scriptsize
\def\arraystretch{1.4}
\centering
\caption{The configurations of RIPE tests.}
\label{tab:ripeconfig}
    \begin{tabular}{c|ccccc}
    \hlinewd{0.6pt}
\multirow{2}{*}{\textbf{Dimensions}} & \multirow{2}{1.5cm}{\centering \textbf{Overflow Technique}} & \multirow{2}{1.5cm}{\centering \textbf{Attack Code}} & \multirow{2}{*}{\textbf{Target Code Pointer}} & \multirow{2}{*}{\textbf{Location}} & \multirow{2}{1cm}{\centering \textbf{Function Abused}} \\
 & & &  &  &\\\hline
\multirow{3}{*}{\textbf{Configurations}} & \multirow{3}{*}{direct, indirect} & \multirow{3}{1.5cm}{\centering rop, createfile, returnintolibc} & \multirow{3}{5cm}{\centering ret, baseptr, funcptrstackvar, funcptrstackparam, structfuncptrstack, funcptrheap, structfuncptrheap, structfuncptrbss, funcptrdata, structfuncptrdata} & \multirow{3}{1cm}{\centering stack, heap, bss, data} & \multirow{3}{1cm}{\centering memcpy, strncpy, strncat}\\
& &  &  & &\\
& &  &  & &\\\hlinewd{0.6pt}
    \end{tabular}
\end{table}

We tested all the valid combinations of the 5 dimensions' configurations, which results in 156 valid attacks in total. No configuration is ignored in each dimension, except ``Function Abused'', which is only for copying data by different functions, and this does not affect the key idea of the attack. Different functions in ``Function Abused'' need different dedicated instrumentation to obtain the store/load addresses and data-lengths as described in Section{~\ref{sec:lib}}, which needs manual design. To avoid too many engineering efforts, we tested three typical functions: {\tt memcpy}, {\tt strncpy}, and {\tt strncat}.

In addition, we prepared a RIPE program without any attack to test if there is any false alarm or not.
It is observed that our DFI system successfully
identifies all the 156 attacks and does not make any false alarm for the case without the attack.

\subsubsection{Heartbleed}
Heartbleed (CVE-2014-0160)~\cite{heartbleed} is a vulnerability in OpenSSL cryptography library. When a message, 
including the payload and the length of the payload, is sent to a server, the server echoes back 
the message with the claimed length. However, it is not checked if the actual payload length is the same
as the claimed one. As such, an attacker may send a message with the actual payload length smaller
than the claimed one. Then, the server sends back not only the original payload but also some additional data, which might be
private sensitive data, to fulfill
the claimed length. Consequently,
sensitive data is stolen by the attacker.
We use the source code in~\cite{heartbleedgit} to simulate such an attack. 
This attack is successfully detected
by our DFI enforcement as the data to be loaded for sending back cannot be most recently written by an instruction not from the sender.
An attack-free transaction, where the actual payload length conforms to the claimed one, is also tested and no false 
alarm is made by our approach.

\subsubsection{Nullhttpd}
Nullhttpd is a HTTP server that has heap overflow vulnerability (CVE-2002-1496) \cite{nullhttpd}. If the server receives a POST request with negative content length $L$, it should not process the request. 
However, the server continues to process and allocates a buffer of $L+1024$ bytes, which is less than 1024 bytes. 
Later, the server writes data of 1024 bytes into the buffer, and therefore buffer overflow occurs. 
The experiment shows that our method successfully detects such buffer overflow. 
When some \texttt{load} instruction attempts to access the data written by 
overflow, it is found that the data is not written by any instructions in the RDS of the \texttt{load} instruction.
An experiment is also conducted to confirm that our approach does not produce false alarm in this context.

\subsubsection{Comparison with HDFI and TMDFI}
To compare the security between our approach and Hardware-assisted Data-Flow Isolation (HDFI)~\cite{Song16},
 we exhaustively tested different tag schemes of HDFI for the example of Fig.~\ref{fig:hdfi_ex},
which are listed in the left three columns of Table~\ref{tab:se_hdfi}.
For each tag scheme, there is some overflow
that cannot be detected by HDFI as shown in column 4,
where \texttt{u0} $\Rightarrow$ \texttt{u1} means some data of \texttt{user0} is 
written into \texttt{user1}'s space through overflow. By contrast, our approach can 
successfully detect all these overflows. 

\begin{table}[!htb]
\scriptsize
\def\arraystretch{1.4}
\centering
\caption{Scenarios for Fig.~\ref{fig:hdfi_ex} where HDFI fails.}
\label{tab:se_hdfi}
\begin{tabular}{ccc|c|c}
\hlinewd{0.6pt}
\multicolumn{4}{c|}{\textbf{HDFI}} & \textbf{Our approach} \\ \cline{1-5}
\texttt{u0}  & \texttt{u1} & \texttt{u2} & Missed overflow & detect?\\\hline
Tag 0 & Tag 0 & Tag 0 & \texttt{u0} $\Rightarrow$ \texttt{u1}, \texttt{u0} $\Rightarrow$ \texttt{u2},  \texttt{u1} $\Rightarrow$ \texttt{u2} & Yes\\
Tag 0 & Tag 0 & Tag 1 & \texttt{u0} $\Rightarrow$ \texttt{u1} & Yes\\
Tag 0 & Tag 1 & Tag 0 & \texttt{u0} $\Rightarrow$ \texttt{u2} & Yes\\
Tag 0 & Tag 1 & Tag 1 & \texttt{u1} $\Rightarrow$ \texttt{u2} & Yes\\
Tag 1 & Tag 0 & Tag 0 &  \texttt{u1} $\Rightarrow$ \texttt{u2} & Yes\\
Tag 1 & Tag 0 & Tag 1 &  \texttt{u0} $\Rightarrow$ \texttt{u2} & Yes\\
Tag 1 & Tag 1 & Tag 0 &  \texttt{u0} $\Rightarrow$ \texttt{u1} & Yes\\
Tag 1 & Tag 1 & Tag 1 & \texttt{u0} $\Rightarrow$ \texttt{u1}, \texttt{u0} $\Rightarrow$ \texttt{u2},  \texttt{u1} $\Rightarrow$ \texttt{u2}  & Yes\\\hlinewd{0.6pt}
\end{tabular}
\end{table}

\begin{figure}[!hbt]
	\centering
	\includegraphics[width=0.6\textwidth]{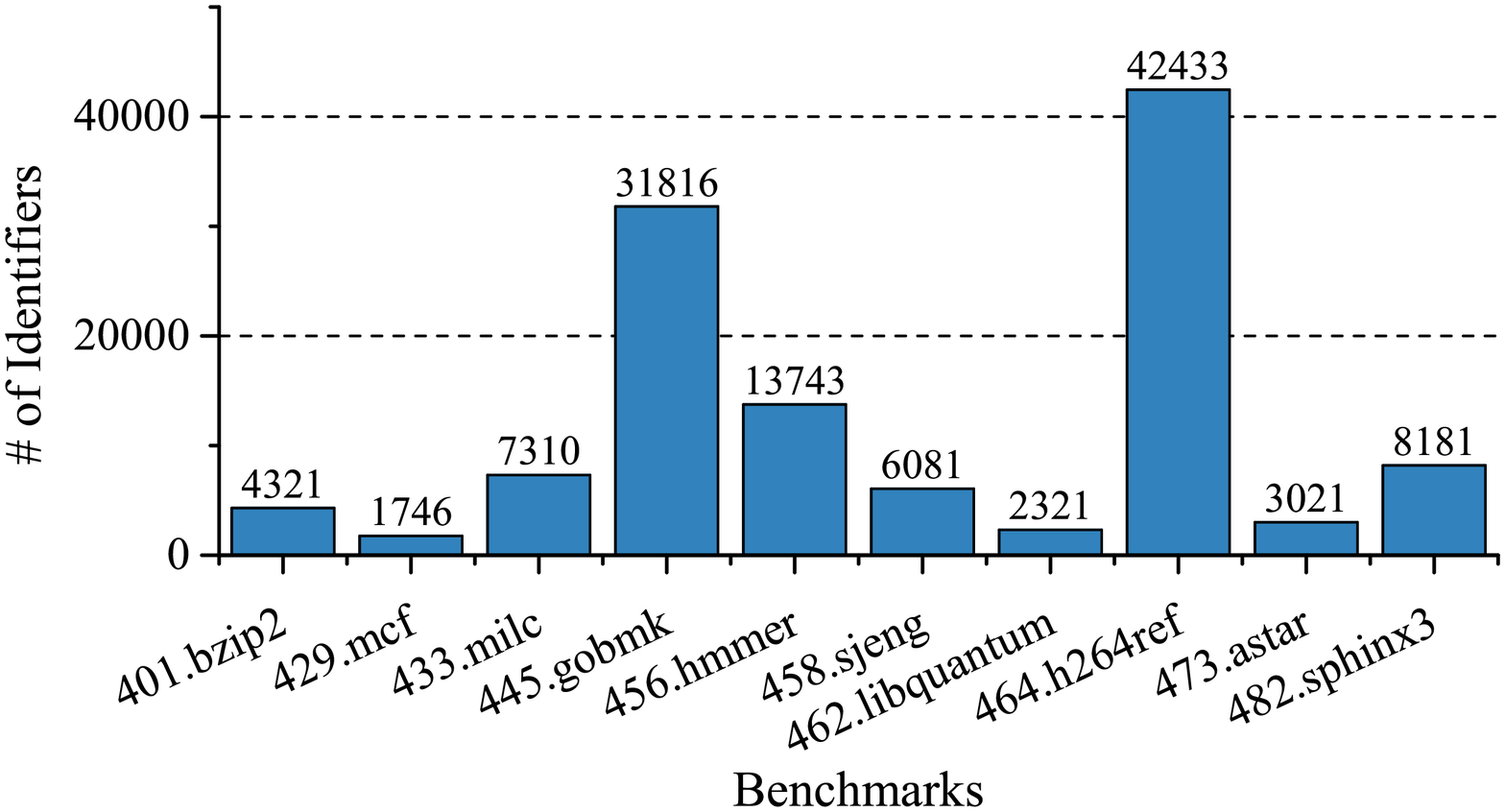}
\vspace{-2ex}
	\caption{The number of identifiers of each benchmark.}
	\label{fig:re_ids}
\vspace{-2ex}
\end{figure}

For TMDFI~\cite{Liu18_2}, although there is a significant improvement over HDFI, its enforcement resolution is still
far from enough in many applications. 
Fig.~\ref{fig:re_ids} shows the numbers of identifiers needed for several benchmarks, which are hundreds or tens of hundreds. Hence, the gap between 
the 256 regions by TMDFI~\cite{Liu18_2} and the actual needs is large. By contrast, our approach can accommodate 
all identifiers in these benchmarks and achieve complete DFI with an overhead similar to 
TMDFI.

In conclusion, although both HDFI and TMDFI are able to identify the attacks in RIPE benchmark, Heartbleed attack, and the attack to Nullhttpd, the data can only be separated into the limited regions. Therefore, for a typical real world program, it is possible that the attacks performed in the same region can pass the checking of HDFI and TMDFI, while our approach can defend against.

\subsection{Performance Overhead}
\label{sec:overhead}

Performance overheads of the following methods are evaluated through
simulations on the SPEC CPU 2006 benchmark~\cite{spec}. 
\begin{itemize}[itemsep=0ex,leftmargin=1em]
    \item Software. This is the original software DFI by \cite{Castro06}.
    \item HBM. This is similar to \cite{Castro06} except that High Bandwidth Memory~\cite{Lee15,Jun17} is employed. 
    \item CMP. This is a parallel approach, where DFI enforcement is performed in another core in CMP with two versions: the software version {\bf CMP-S} (multithreading) and the hardware version {\bf CMP-H} using our info-collector circuit.
    \item PIM. This is the proposed hardware-assisted parallel approach using PIM.
\end{itemize}
Our {\bf proposed approach} has two variants: CMP-H and PIM. 
As architectural simulations using gem5 are many orders of magnitude slower than real system runs for accurately modeling hardware behaviors, we manage to terminate the simulations to run the region of interest of the program for sufficiently long time while accounting for a reasonable simulation time.
To ensure a fair comparison, 
each application was terminated at the same point in the simulations.
The results are summarized in Table~\ref{tab:overhead}. 

\begin{table*}[!htb]
\def\arraystretch{1.5}
\centering
\caption{Performance overhead of DFI. $^\dagger$Computation time of optimizations and compression is neglected. $^\ddagger$Computation time of optimizations and compression is considered. $^\mathsection$No DFI packet is sent to the memory.}
\label{tab:overhead}
\resizebox{1.0\textwidth}{!}{
\begin{tabular}{c|c|c|c|cc|cccc|cc|ccc}
\hlinewd{1pt}
\multicolumn{2}{c|}{}  & \multicolumn{1}{c|}{\textbf{Software}~\cite{Castro06}} & \textbf{HBM} & \textbf{CMP-S} & \textbf{CMP-H} & \multicolumn{4}{c|}{\textbf{PIM (No Compression or Optimization)}} & \multicolumn{2}{c|}{\textbf{PIM (512B Buffer)}} & \multicolumn{3}{c}{\textbf{PIM (2KB Buffer)}} \\\hline
\multicolumn{2}{c|}{\textbf{Column ID}} & 1 & 2 & 3 & 4 & 5 & 6$^\mathsection$ & 7 & 8 & 9 & 10 & 11 & 12 & 13 \\\hline
\multicolumn{2}{c|}{\textbf{Compression}} & $\times$ & $\times$ & $\times$ & $\surd$ & $\times$ & $\times$ & $\surd$ & $\times$ & $\surd$ & $\surd$ & $\surd$ & $\surd$ & $\surd$ \\
\multicolumn{2}{c|}{\textbf{Transmit Buf Size}} & - & - & - & 2KB & - & - & 2KB & 2KB & 512B & 512B& 2KB & 2KB & 2KB \\
\multicolumn{2}{c|}{\textbf{Runtime Optimization}} & $\times$ & $\times$ & $\times$ & All & $\times$ & $\times$ & E & A,B,C,D & All & C,E & All  & C,E & C,E \\
\multicolumn{2}{c|}{\textbf{\#Gates in Info-Collector}} & - & - & - & $^\dagger$ & $<$2908$^\dagger$ & $^\dagger$ & $^\dagger$ & $^\dagger$ & $^\dagger$ & 116,769$^\ddagger$  & $^\dagger$ & $^\dagger$ & 753,666$^\ddagger$  \\\hline
\multirow{10}{*}{\rotatebox[origin=c]{90}{\textbf{Benchmark}}}  & \textit{401.bzip2} & 218.7\% & 219.5\% & 543.3\% & 43.7\% & 313.4\% & 34.6\% & 40.0\% & 44.7\% & 40.8\% & 43.2\% & 37.9\% & 38.6\% & 39.8\% \\
 & \textit{429.mcf} & 105.0\% & 105.6\% & 320.5\% & 28.8\% & 191.6\% & 18.9\% & 24.3\% & 27.1\% & 25.7\% & 26.8\% & 23.9\% & 24.1\% & 24.8\% \\
 & \textit{433.milc} & 80.9\% & 82.7\% & 256.6\% & 24.1\% & 150.0\% & 22.5\% & 25.5\% & 24.1\% & 25.3\% & 26.6\% & 23.4\% & 24.4\% & 25.0\% \\
 & \textit{445.gobmk} & 179.0\% & 179.0\% & 463.0\% & 59.4\% & 272.3\% & 46.9\% & 54.8\% & 56.5\% & 55.9\% & 57.7\% & 53.5\% & 54.3\% & 55.3\% \\
 & \textit{456.hmmer} & 233.4\% & 233.5\% & 1087.6\% & 60.9\% & 510.7\% & 47.2\% & 55.5\% & 64.2\% & 57.9\% & 60.8\% & 53.0\% & 53.0\% & 55.0\% \\
 & \textit{458.sjeng} & 372.6\% & 374.2\% & 226.9\% & 29.4\% & 128.6\% & 24.6\% & 28.0\% & 28.5\% & 28.6\% & 29.4\% & 27.3\% & 27.6\% & 28.2\% \\
 & \textit{462.libquantum} & 61.2\% & 61.2\% & 262.2\% & 22.5\% & 156.4\% & 21.9\% & 23.9\% & 23.2\% & 24.2\% & 25.0\% & 22.6\% & 22.7\% & 23.3\% \\
 & \textit{464.h264ref} & 205.3\% & 205.8\% & 544.0\% & 44.5\% & 275.7\% & 33.9\% & 42.1\% & 42.4\% & 42.8\% & 45.2\% & 39.9\% & 41.0\% & 43.1\% \\
 & \textit{473.astar} & 116.6\% & 116.6\% & 442.0\% & 38.2\% & 255.7\% & 31.6\% & 36.9\% & 38.4\% & 37.2\% & 39.1\% & 35.2\% & 35.5\% & 36.5\% \\
 & \textit{482.sphinx3} & 41.4\% & 41.6\% & 123.0\% & 18.7\% & 74.4\% & 32.1\% & 33.4\% & 33.4\% & 33.6\% & 33.9\% & 33.1\% & 33.1\% & 33.3\% \\\hline
  \multicolumn{2}{c|}{\textbf{Average}} & \textbf{161.4\%} & \textbf{162.0\%} & \textbf{426.9\%} & \textbf{37.0\%} & \textbf{232.9\%} & \textbf{31.4\%} & \textbf{36.4\%} & \textbf{38.2\%} & \textbf{37.2\%} & \textbf{38.8\%} & \textbf{35.0\%} & \textbf{35.4\%} & \textbf{36.4\%}\\\hlinewd{1pt}

\end{tabular}
}
\end{table*}

On average, the performance overhead of software DFI{~\cite{Castro06}} is $161\%$ as shown in column 1. 
Column 2 shows the result of software DFI using HBM, where
the memory bandwidth is abundant and memory access latency
is fairly low. One can see that using HBM brings almost 
no overhead reduction. This result confirms the analysis in Section~\ref{sec:sys}. 
The results of the parallel approach using another CMP core 
are summarized in columns 3 and 4, for software and our hardware version, respectively. Without dedicated hardware, the parallel approach actually increases the overhead due to the expensive communication in software. CMP-H reduces the overhead to $37\%$.

The PIM results are listed in columns 5-13, where ``All'' means all of the 5 optimization techniques are applied and ``C, E'' corresponds to the results where only the two most effective optimizations are employed. In column 5, the overhead is
$233\%$ although the offline optimization has been applied.
This tells the importance of our hardware-based optimization and compression. In column 6, we dropped all DFI packets without sending them out by simulating only instruction fetching but not executions of instrumentation. This is not realistic for DFI, but is to obtain a lower bound for the overhead, which is about $31\%$. Column 7 shows that the joint effect of data compression and optimization E is dramatic. Please note optimization E
is designed for increasing the chance of data compression.
The setup for column 11 is very similar to column 4, 
except that one is by PIM and the other is by CMP.
Examining the results of the two columns that their overhead reductions are similar. PIM is a little better as it causes
less cache contentions as CMP. Column 13 takes the two most important optimizations and considers the compression/optimization delay, showing an overhead of about $36\%$.

\begin{figure}[!hbt]
	\centering
	\includegraphics[width=0.4\textwidth]{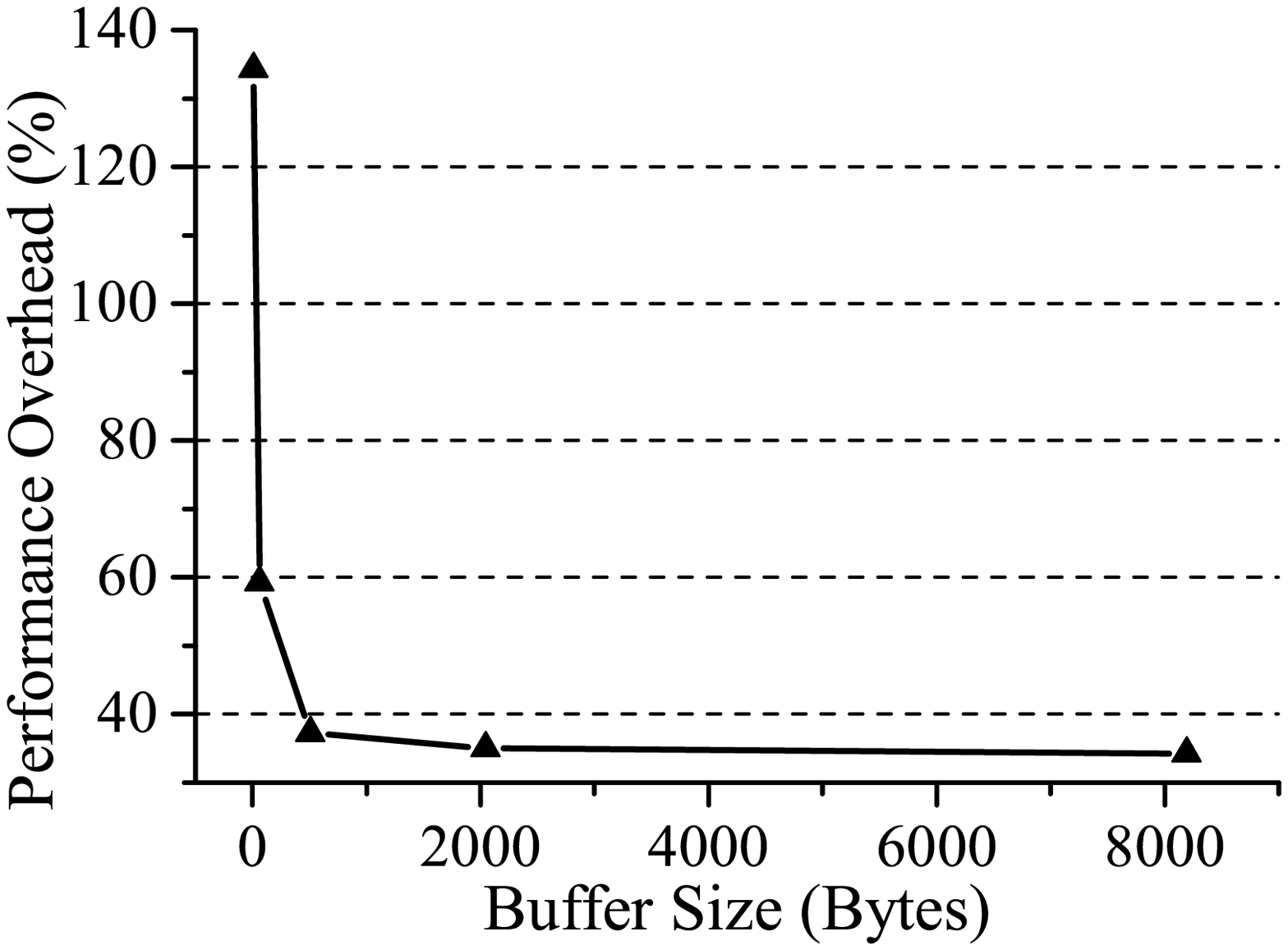}
	\caption{Overhead vs. buffer size.}
	\label{fig:re_opt_buf}
\end{figure}

\begin{nohyphen}
The effect of transmission buffer size on reducing
performance overhead is plotted in Fig.~\ref{fig:re_opt_buf}. 
It shows that an increase of buffer size from 0 quickly brings down the overhead. However, the reduction soon diminishes as buffer size reaches 2K bytes and
this is why we limit the buffer size to be no more than 2K in our experiments. 


\end{nohyphen}

\begin{figure}[!hbt]
	\centering
	\includegraphics[width=0.7\textwidth]{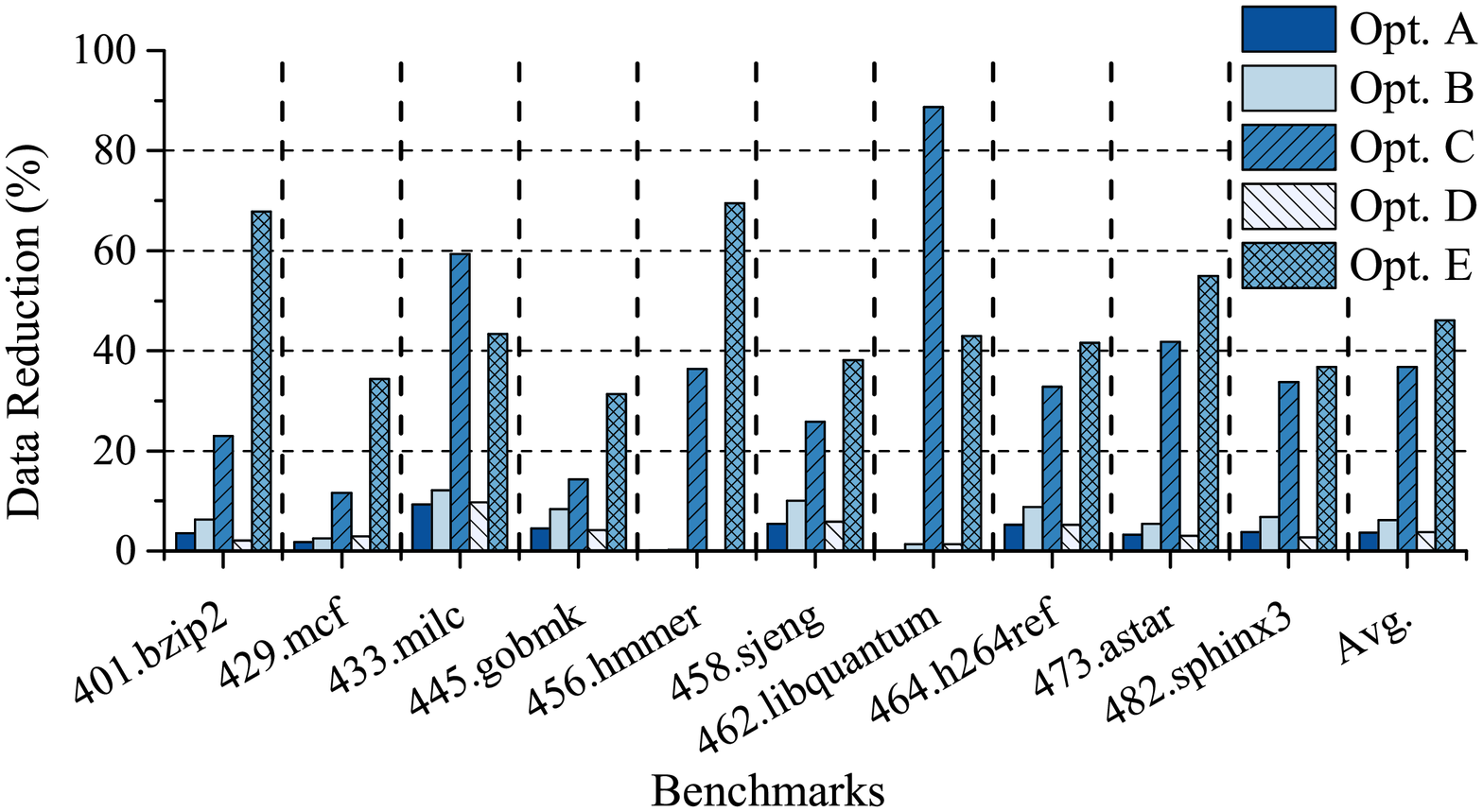}
	\caption{Effects of optimization techniques.}
	\label{fig:re_opt}
\end{figure}

The effects of the 5 optimization techniques described in Section~\ref{sec:dfi_buf} on data reduction are evaluated
separately and the results are depicted in Fig.~\ref{fig:re_opt}. It shows that optimizations C and E always lead to more data reduction than
the other techniques. For \textit{462.libquantum}, optimization C can reduce data by over $80\%$ while optimization E reduces data by more than $60\%$ for both
\textit{401.bzip2} and \textit{456.hmmer}. Optimization E is designed to facilitate compression, and one can observe that its average data reduction is $46\%$, which is also the average {\bf compression ratio}.

\subsection{Tradeoff Between Detection Latency and Overhead}

\begin{figure}[!hbt]
	\centering
	\includegraphics[width=0.45\textwidth]{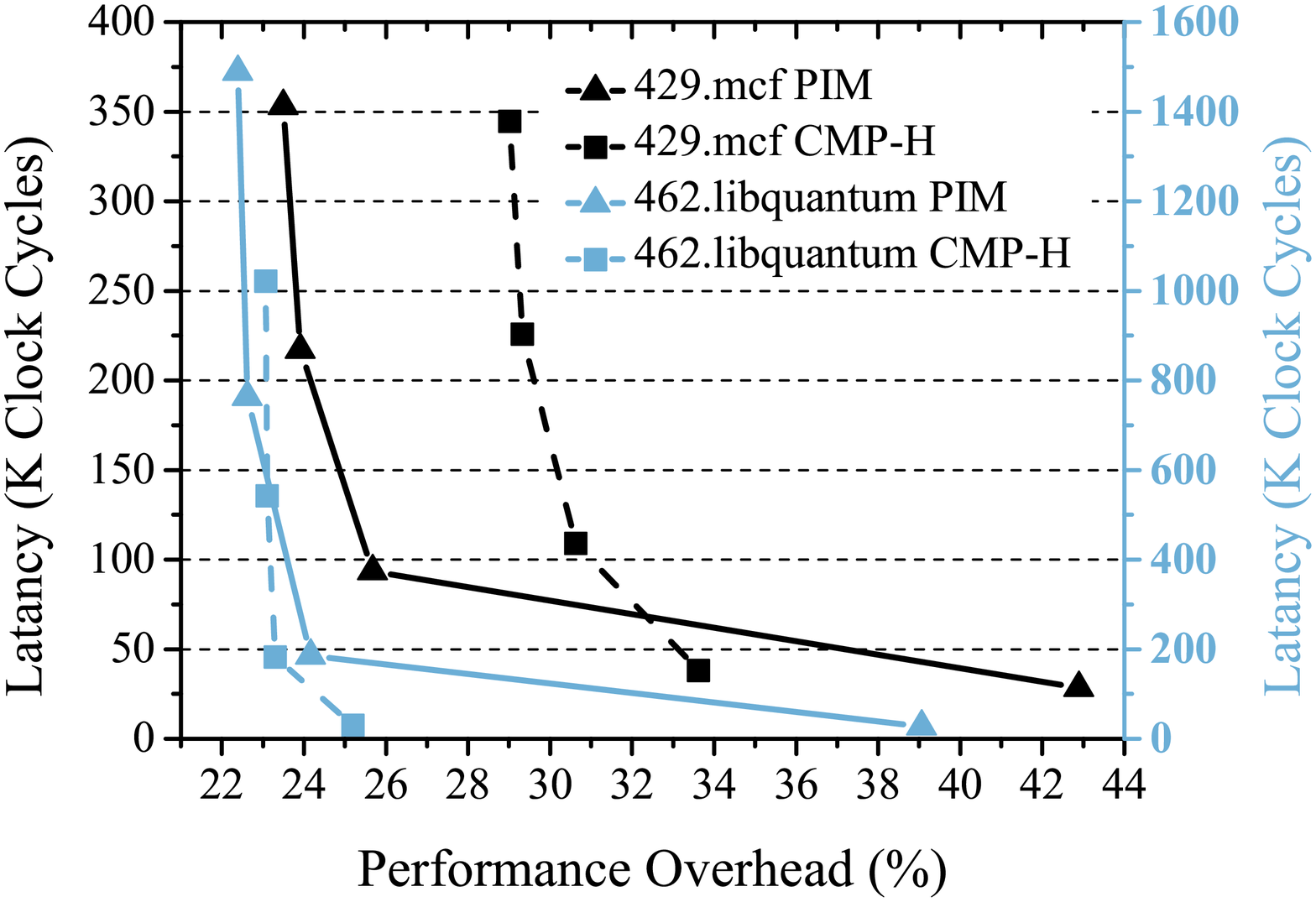}
	\caption{Detection latency vs. overhead for \textit{429.mcf} and \textit{462.libquantum}.}
	\label{fig:re_ptg_ov}
\end{figure}

Ideally, the latency for detecting DFI violations need to be minimized so that attackers have less time to complete 
damaging operations. In Fig.~\ref{fig:re_ptg_ov}, we show that 
the latency can be managed by a tradeoff with the overhead via
varying the buffer size.  The results also indicate that the PIM approach performs better for low overhead while the CMP-H approach is slightly better for obtaining low latency. 
The reason is, as discussed in Section{~\ref{sec:overhead}}, PIM causes fewer cache contentions as CMP, and leads to less overhead. However, due to the data transferring latency from the main processor to the PIM and the relatively low performance of PIM, using PIM has longer latency than using CMP.

\subsection{Binary Size and Hardware Circuit Overhead}

\begin{figure}[!hbt]
	\centering
	\includegraphics[width=0.5\textwidth]{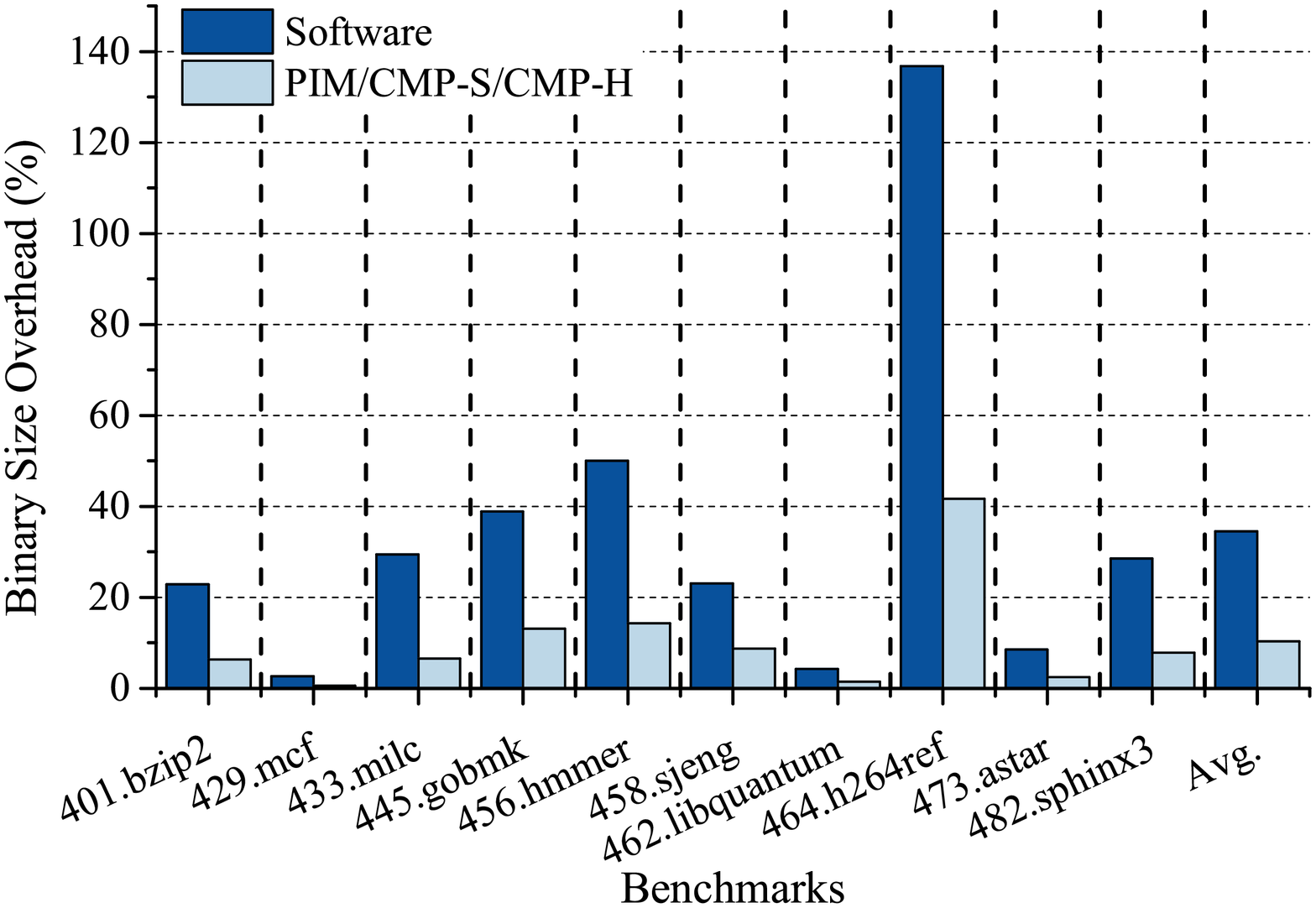}
	\caption{The executable binary size overhead of different benchmarks.}
	\label{fig:binov}
\end{figure}

Performing instrumentation can increase the size of the executable binary of the user program.
As shown in Fig.{~\ref{fig:binov}}, our approach only increases the binary size by 10\% on average, while software-DFI can increase the size by 35\%, which is $3\times$ as our approach. Note that the amounts of the instrumentation are the same for PIM, CMP-S and CMP-H.

The info-collector circuit is implemented by synthesizing Verilog using Synopsys Design Compiler and ASAP 7nm cell library~\cite{asap}. The info-collector 
with basic operation and compression costs only 2908 gates and less than 30ps circuit delay. Hence, its area and delay are negligible. 
We also implemented the circuit for optimization C/E. The results with these implementations are in columns 10 and 13 of Table~\ref{tab:overhead}, where
the gate counts of the info-collector with different buffer sizes are listed.
The hardware circuit overhead is dominated by the optimization part. The gate count of 754K is not trivial, but still a small fraction of a modern microprocessor 
that often has hundreds of millions of gates. 
Moreover, our DFI can isolate data among 64K regions and the hardware cost per region is no more than 12 gates. 
Although the hardware overhead of our approach is still high for practical use of many current civil applications, there can be niche applications, where security is super critical while hardware cost is of little concern, such as some military applications. Moreover, as security problems become increasingly prevalent and hardware cost continues to decrease, there can be a point in future where the hardware cost becomes justified for the achieved security.
The works of CHERI{~\cite{Watson15}} and HDFI{~\cite{Song16}} did not describe their hardware details.
However, HDFI can isolate only between 2 regions, and its hardware cost is almost impossible to be less than 24 gates. Therefore, it is highly possible that the hardware cost per region of our approach is less than HDFI.

The memory overhead of our approach is $N$/$2$ and the size of the RDSs if the user program needs $N$ bytes data memory, and CHERI and HDFI have much less memory overhead than us. In our design, we trade more memory space for higher security.

\section{Discussions}
\label{sec:limit}

In this section, we discuss the trade-off and limitation of the asynchronous enforcement, as well as potential supports to OS. 

\textbf{The Asynchronous Enforcement:} Different from other works such as the software DFI{~\cite{Castro06}}, CHERI{~\cite{Watson15}} and HDFI{~\cite{Song16}}, of which the DFI enforcement approaches are synchronous, our approach is asynchronous and has detection latency.
The security risk due to asynchronous checking is a price paid for the performance overhead reduction compared to the software DFI. On the other hand, the proposed approach arguably still provides a higher level of security than CHERI and HDFI, where a large amount of attacks are completely undefended. Specifically, if an attacker violates DFI for two data in one region for HDFI, this violation cannot be detected by HDFI. Although our asynchronous check leaves a brief window before stopping the program execution, any followup attacks must be carried out within this window and the threshold for such attacks is significantly raised. In contrast, in many cases, attackers can launch attacks without such restrictions for systems with HDFI.

\textbf{The Supports to the OS:} User programs and OS are two main application scenarios of DFI. Our main focus is the principle techniques for reducing DFI overhead and working out the details for user programs. The same principles are applicable for OS, but the implementation details are quite different, which is out of the scope of this work.

\begin{sloppypar}
\section{Conclusions and Future Research}
\label{sec:conclu}
\end{sloppypar}

Data-Flow Integrity (DFI) is potentially a very powerful security measure that
can detect a large number of software attacks. However, it requires checking
a large volume of data and thus intrinsically entails huge performance overhead.
We propose a hardware-assisted parallel approach to address
this challenge.
This approach can reduce the overhead by more than $4\times$ compared
to the original software DFI while enforcing complete DFI. 
In future research, we will study how to further reduce the performance overhead 
and detection latency.


\bibliographystyle{ACM-Reference-Format}
\bibliography{ref}


\end{document}